\documentclass[twocolumn,fleqn]{svjour2}    % twocolumn
\usepackage{graphicx}
\usepackage{graphicx}% Include figure files
\usepackage{dcolumn}% Align table columns on decimal point
\usepackage{bm}% bold math
\usepackage{subfigure}% sub figure
\usepackage{amsmath}
\usepackage{graphicx}
\def\gsim{\;\lower4pt\hbox{${\buildrel\displaystyle >\over\sim}$}\;}
\def\lsim{\;\lower4pt\hbox{${\buildrel\displaystyle <\over\sim}$}\;}
\def\grls{\;\lower4pt\hbox{${\buildrel\displaystyle >\over <}$}\;}
\journalname{Astrophysics and Space Science}
\begin{document}

\title{Dynamic Evolution of a Quasi-Spherical
General Polytropic Magnetofluid with Self-Gravity }

\author{Wei-Gang Wang        \and
        Yu-Qing Lou
}


\institute{W.-G. Wang \at
              Physics Department and Tsinghua Center
for Astrophysics (THCA), Tsinghua University, Beijing 100084,
China;
          \email{wwg03@mails.tsinghua.edu.cn; weigwang@stanford.edu}
          \and
           Y.-Q. Lou \at
           1. Physics Department and the Tsinghua Center for
Astrophysics (THCA), Tsinghua University, Beijing 100084, China;\\
2. Department of Astronomy and Astrophysics, The University
of Chicago, 5640 S. Ellis Avenue, Chicago, IL 60637 USA;\\
3. National Astronomical Observatories, Chinese Academy
of Sciences, A20, Datun Road, Beijing 100012, China.\\
\email{louyq@tsinghua.edu.cn; lou@oddjob.uchicago.edu} }

\date{Received: date / Accepted: date}
\maketitle

\begin{abstract}
In various astrophysical contexts, we analyze self-similar
behaviours of magnetohydrodynamic (MHD) evolution of a
quasi-spherical polytropic magnetized gas under self-gravity with
the specific entropy conserved along streamlines. In particular,
this MHD model analysis frees the scaling parameter $n$ in the
conventional polytropic self-similar transformation from the
constraint of $n+\gamma=2$ with $\gamma$ being the polytropic
index and therefore substantially generalizes earlier analysis
results on polytropic gas dynamics that has a constant specific
entropy everywhere in space at all time. On the basis of the
self-similar nonlinear MHD ordinary differential equations, we
examine behaviours of the magnetosonic critical curves, the MHD
shock conditions, and various asymptotic solutions. We then
construct global semi-complete self-similar MHD solutions using a
combination of analytical and numerical means and indicate
plausible astrophysical applications of these magnetized flow
solutions with or without MHD shocks. \keywords{clusters of
galaxies \and magnetohydrodynamics
%stars: AGB and post-AGB
\and stars: formation \and stars: neutron \and stars:
winds, outflows \and supernovae: general}
\PACS{95.30.Qd\and98.38.Ly\and95.10.Bt\and97.10.Me\and97.60.Bw}
\end{abstract}

%width=0.75\textwidth
\section[]{Introduction}\label{intro}
\renewcommand{\topfraction}{0.98}

Dynamic evolutions of various astrophysical shock flows under
gravity on completely different spatial and temporal scales have
been attracting astrophysicists, mainly because of a rich variety
of observational phenomenology in different astrophysical systems,
such as stellar collapses, supernova explosions and supernova
remnants (e.g., Bethe et al. 1979; Goldreich \& Weber 1980;
Chevalier 1982; Yahil 1983; Lattimer \& Prakash 2004; Lou \& Wang
2006, 2007; Lou \& Cao 2008; Lou \& Hu 2008 in preparation), star
formation, proto-stellar core formation as well as `champagne
flows' in star-forming molecular clouds (e.g., Shu 1977; Shu et
al. 1987; Tsai \& Shu 1995; Shu et al. 2002; Shen \& Lou 2004;
Bian \& Lou 2005; Lou \& Gao 2006), formation and spherical
accretions of black holes (e.g., Bondi 1952; Bahcall \& Ostriker
1975; Hennawi \& Ostriker 2002; Cai \& Shu 2005; Hu et al. 2006),
active galactic nuclei (AGNs) at the galactic level (e.g., Small
\& Blandford 1992) and the formation of galaxy clusters and voids
in the Universe (e.g., Gunn \& Gott 1972; Fillmore \& Goldreich
1984a, b; Bertschinger 1985) as well as dynamic galaxy cluster
winds (Lou, Jiang \& Jin 2008; Jiang \& Lou 2008 in preparation).
This broad class of theoretical models poses challenges for a
deeper and simpler theoretical understanding of relevant
astrophysical shock flows under consideration. Meanwhile, these
problems also involve a diverse range of physical processes and
the relevant model investigations often involve numerical
hydrodynamic simulations with data input from high-energy physics
experiments and/or theories (e.g., Janka \& M\"uller 1996 for
supernovae).

In this paper, we mainly focus on basic hydrodynamic and
magnetohydrodynamic (MHD) aspects of this class of problems
because of the relative simplicity and generality, allowing for
various novel and interesting solution features to manifest. We
introduce the general polytropic description to subsume several
unspecified energetic processes into a few index parameters. There
might be certain evolution phases where a simple and direct
analysis works almost as well as numerical simulations do (e.g.,
Janka \& M\"uller 1996), and in this situation such an analysis
would be extremely valuable for physical insight (e.g., Yahil
1983) and for testing numerical codes in construction. In this
spirit, we advance a theoretical MHD flow model with this kind of
simplification (e.g., self-similarity) and idealization (e.g., a
completely random transverse magnetic field and specific entropy
conservation along streamlines) to reveal plausible MHD behaviours
and shocks of magnetized gas flows of astrophysical interests.

While self-similar evolutions of gas dynamics in spherical
geometry have been actively pursued early on (e.g., Sedov 1959;
Bodenheimer \& Sweigart 1968; Larson 1969a, b; Penston 1969a, b),
extensively analyzed by many in diverse astrophysical contexts
(e.g., Shu 1977; Hunter 1977; Cheng 1978; Goldreich \& Weber 1980;
Chevalier 1982; Yahil 1983; Whitworth \& Summers 1985; Tsai \& Hsu
1995; Shu et al. 2002; Shen \& Lou 2004; Fatuzzo et al.
%, Adams \& Myers
2004; Lou \& Gao 2006) and significantly reformulated in various
physical aspects (e.g., Terebey, Shu \& Cassen 1984; Suto \& Silk
1988; Chiueh \& Chou 1994; McLaughlin \& Putritz 1997; Boily \&
Lynden-Bell 1995; Cai \& Shu 2005), we have found novel features
and generalized similarity solutions for this type of classical
nonlinear problems in recent years. As examples, our new
similarity solution features include envelope expansion with core
collapse (EECC) solutions which smoothly pass through the sonic
critical line twice (Lou \& Shen 2004; Lou \& Gao 2006),
quasi-static asymptotic solutions (Lou \& Wang 2006), quasi-static
asymptotic MHD solutions (Lou \& Wang 2007; Wang \& Lou 2007),
quasi-static asymptotic solutions in two gravity-coupled fluids
(Lou, Jiang \& Jin 2008; Jiang \& Lou 2008 in preparation), strong
magnetic field solutions (Yu \& Lou 2006; Wang \& Lou 2007), and
various new (MHD) shock solutions (Shen \& Lou 2004; Bian \& Lou
2005; Yu et al.
%, Lou, Bian \& Wu
2007).

In addition to the known Larson-Penston (LP) type solutions and
central free-fall solution (Shu 1977), the existence of several
new nonlinear self-similar MHD polytropic flow solutions reveals
multiple clues and leads to new concepts.

First of all, knowing these self-similar nonlinear solutions, we
realize that a number of self-similar behaviours may possibly
occur in seemingly similar flow systems, and the evolution problem
of `which goes where' remains an important open question. Clearly,
this cannot be immediately answered within the self-similar
framework. While it is generally thought that self-similar
behaviour gradually emerges when a dynamical system evolves
sufficiently far away from its initial and boundary conditions, we
would emphasize that initial and boundary conditions do affect the
nonlinear self-similar behaviour in the sense of which similarity
behaviour emerges eventually. In other words, the final
self-similar phase still retains memory of a certain class of
initial and boundary conditions. As an example, Tsai \& Hsu (1995)
introduced a self-similar behaviour that has the outer (initial)
configuration the same as that of Shu (1977), but their inner
(final) behaviour is qualitatively different from that of Shu
(1977): the model of Tsai \& Hsu (1995) includes a self-similar
isothermal shock and also the central mass accretion rate differs
from that of Shu (1977). As suggested by simulation results of
Tsai \& Hsu (1995), a stronger push at the centre of a static
isothermal sphere may result in a shock, while a weaker push will
lead to an isothermal expansion-wave collapse solution (EWCS; see
their Figs. $1-3$).
%{\bf A discussion of Tsai \& Hsu (1995) as an example
%to make the point more concrete.} {\it To YQL: done.}

Secondly, although former nonlinear self-similar solutions were
often obtained from numerical simulations first (e.g., Larson
1969a, b; Penston 1969a, b), the very existence of different
self-similar behaviours in the same physical setting would imply
that so far numerical hydrodynamic simulations might have not been
sufficiently thorough in exploring the overall dynamic evolution
features of nonlinear flows with or without shocks. These
semi-analytical solutions are important clues and benchmarks for
numerical code development. More importantly, numerical
simulations can further tell whether new self-similarity solutions
obtained in our analysis can be actually realized under realistic
flow situations, by introducing different initial and boundary
conditions for a similar flow system and by observing which
similarity behaviour does it manifest eventually.

Finally, it is also possible that one flow system may not behave
self-similarly in the strict sense, nonetheless its evolution can
be qualitatively explained by intuitions gained by examining
various self-similar solutions: for example, the central free-fall
phase of Shu (1977) is common for a wide range of polytropic index
$\gamma$; the existence of other new solutions requires
specialized conditions. Physically, our new asymptotic solutions
have different features in terms of the balance and competition of
the involved forces, which tells us about the possible asymptotic
flow behaviours.
%{\bf Should provide a better introduction}

As an essential ingredient in our theoretical model development,
magnetic fields are important in various astrophysical settings;
their presence in ionized plasma systems is ubiquitous and they
play dynamical as well as diagnostic roles in many ways; they are
responsible for sporadic violent activities and for producing
relativistic particles such as cosmic rays through shocks and
reconnection. The generation of magnetic fields is related to
convective motions, sustained turbulence and differential
rotations, via dynamo effects or magnetorotational instabilities
(MRI) (e.g., Parker 1979; Thompson \& Duncan 1993 for dynamo
effects of magnetized pulsars; Chandrasekhar 1961; Balbus \&
Hawley 1998; Balbus 2003 for disc magnetorotational
instabilities).
%{\bf should learn and mention key references of
%Velikhov-Chandrasekhar-Balbus-Hawley instability;
%e.g., Chandrasekhar 1961; Balbus \& Hawley 1998?
%Rev Mod Phys.; Balbus 2003}{\it to YQL: done.} ).
For the purpose of our formulation here, we take the presence of
magnetic field for granted and simply presume that a completely
random magnetic field permeates in a gas medium with self-gravity.
This gas medium can be stellar interior, hot stellar coronae,
interstellar medium (ISM), or intracluster medium (ICM) and so
forth.

During the collapse phase of a quasi-spherical system under
self-gravity, a highly conducting magnetofluid undergoes a
magnetic field enhancement through the magnetic flux conservation,
commonly referred to as the frozen-in condition for magnetic
field. When rotation is sufficiently slow in an astrophysical
system, the overall geometry may remain quasi-spherical and our
model analysis with the quasi-spherical random-field approximation
(e.g., Zel'dovich \& Novikov 1971) can be applicable
%The rotational generation mechanisms are supposed to determine
%the magnetic field strength, yet The quasi-spherical random-field
%approximation (e.g., Zel'dovich \& Novikov 1971) takes care of the
%large-scale evolution of this generated random magnetic field
(Yu \& Lou 2005; Yu et al. 2006; Lou \& Wang 2007; Wang \& Lou
2007; Lou \& Hu 2008, in preparation). Chiueh \& Chou (1994) first
studied a quasi-spherical isothermal MHD problem by including the
magnetic pressure gradient into the radial momentum equation.
However,
%A further analysis on the magnetic force reveals that
the magnetic tension force does not average to zero (Yu \& Lou
2005; Wang \& Lou 2007) and thus should be also included. Although
on small scales, magnetic tension force tends to drive
inhomogeneous gas motions, we presume that these non-spherical
flows may be neglected as compared to the large-scale radial bulk
motion of gas. The key point is that the (self-)gravity is strong
enough to hold on the entire gas mass and induce core collapse.
Hence on large scales in a quasi-spherical geometry, a completely
random magnetic field contributes to the dynamics in the form of
the average magnetic pressure gradient force and the average
magnetic tension force in the radial direction.
%Considering the formidable work needed in describing the magnetic
%force, especially tension force, our simplification and idealization
%intend to demonstrate relevant dynamic effects of the magnetic field
%in an intuitive and direct way.

In a series of papers during past several years, we have
systematically investigated isothermal MHD problems (Yu \& Lou
2005; Yu et al. 2006) and polytropic MHD problems in the special
case of a {\it constant} specific entropy in space at all time
(i.e., $p=\kappa\rho^\gamma$ with a constant $\kappa$, a pressure
$p$, a mass density $\rho$, and a polytropic index $\gamma$; Wang
\& Lou 2007; Lou \& Wang 2007; Lou \& Hu 2008, in preparation).
Under certain circumstances, the isothermality may be a crude yet
sensible first approximation for astrophysical fluid dynamics with
or without magnetic field (e.g., for star formation processes
within a collapsing molecular cloud, Shu 1977; hot bubbles and
superbubbles in ISM produced by supernovae, Lou \& Zhai 2008, in
preparation; or central core collapse region of a globular
cluster, Bahcall \& Ostriker 1975; Inagaki \& Lynden-Bell 1983),
even for modeling shocks in dynamical processes (e.g., Courant \&
Friedrichs 1976; Spitzer 1978). In other situations, a polytropic
fluid with a {\it constant} specific entropy is invoked for
theoretical model investigation (e.g., Goldreich \& Weber 1980 for
a stellar collapse of a relativistically hot gas priori to the
rebound process; Yahil 1983 for gravitational core collapses; Suto
\& Silk 1988; Lou \& Cao 2008 for rebound processes in stellar
core collpases). While the polytropic approximation with a
constant specific entropy is a useful description for various
astrophysical processes, there is however no obvious reason why
the entropy should necessarily remain constant in a polytropic
gas. In dynamic processes, a more general situation involves the
`specific entropy' conservation along streamlines; and this
includes the constant entropy as a special case.
%Usually, the conserved quantity is the specific
%entropy of an adiabatic fluid parcel, or say,
%the specific entropy conserves `along streamlines'.
Therefore, self-similar polytropic gas evolution under this
`equation of state' represents a significant generalization of the
`constant entropy' problem, and can be more adaptable to various
dynamic model applications in astrophysics.

Indeed, several authors have adopted such an `equation of state'
and carried out their model analyses to various extents (e.g.,
Cheng 1978; Chevalier 1982; Fatuzzo et al. 2004; Wang \& Lou 2007;
Lou \& Cao 2008; Lou \& Hu 2008, in preparation). In technical
terms, adopting such an `equation of state' is equivalent to
freeing the choice of index parameter $n$ in the usual
self-similar transformation equation (\ref{transform1}) from the
constraint $n+\gamma=2$ with $\gamma$ being the polytropic index
(Wang \& Lou 2007). Here, this index parameter $n$ holds the key
to the scaling of physical variables under consideration, such as
mass density, random magnetic field strength, radial flow
velocity, pressure, and temperature in the outer portion of a
quasi-spherical gas. As $\gamma$ is a thermodynamic parameter, we
expect that $n$ and $\gamma$ parameters should be independent of
each other in general. For instance, Lou \& Wang (2007) have shown
that the radial scaling of a completely random magnetic field is
actually independent of the polytropic index $\gamma$. The main
motivation of this paper is to revisit the self-similar MHD
problem (Wang \& Lou 2007) but with an equation of state for
specific entropy conservation along streamlines and to generalize
our new self-similar hydrodynamic and MHD solutions into a
substantially larger parameter regime.

This paper with special emphasis on theoretical aspects is
structured as follows. Our motivation and background information
are provided in Section 1 as an introduction. Section 2 describes
the basic formalism on nonlinear MHD flows with a completely
random transverse magnetic field, including the self-similar
transformation and MHD shock conditions. Section 3 presents
various analytical asymptotic MHD solutions for large and small
$x$ as well as along the magnetosonic critical curve; these
asymptotic MHD solutions are valuable for understanding relevant
physics and can be utilized to construct global semi-complete
solutions which are valid in the range of $0<x<+\infty$. Section 4
shows examples of various possible semi-complete global solutions
with or without MHD shocks and outlines astrophysical applications
of these solution types. We discuss our results in Section 5.
Mathematical details are included in several appendices for the
convenience of reference.

\section[]{Formulation of the Model Problem}

\subsection[]{Nonlinear Magnetohydrodynamic Equations}
%\section[]{Similarity MHD Flows}\label{mhdformulation}
%\subsection{MHD Formulation of the Problem}\label{form}

In spherical polar coordinates ($r$, $\theta$, $\phi$), the basic
nonlinear equations for a quasi-spherical MHD evolution of a
general polytropic gas under self-gravity include the mass
conservation equation
\begin{equation}\label{basic1}
\frac{\partial\rho}{\partial t}+\frac{1}{r^{2}}
\frac{\partial}{\partial r}(r^{2}\rho u)=0\ ,
\end{equation}

\begin{equation}\label{basic2}
\frac{\partial M}{\partial t}+u\frac{\partial M}{\partial r}=0\ ,
\end{equation}

\begin{equation}\label{basic3}
\frac{\partial M}{\partial r}=4\pi r^{2}\rho\ ,
\end{equation}
where $\rho(r,t)$ is the gas mass density, $u(r,t)$ is the bulk
radial flow speed, $M(r,t)$ is the enclosed mass within radius $r$
at time $t$; the radial momentum equation
\[
\rho\bigg(\frac{\partial u}{\partial t} +u\frac{\partial u}{\partial
r}\bigg)=-\frac{\partial p}{\partial r} -\frac{GM\rho}{r^{2}}
\]
\begin{equation}\label{basic4}
\qquad\qquad\qquad\qquad -\frac{\partial}{\partial r}\bigg(\frac{
<B_{t}^{2}>}{8\pi}\bigg)-\frac{<B_{t}^{2}>}{4\pi r}\ ,
\end{equation}
where $p$ is the thermal gas pressure, $<B_{t}^{2}>$ is the ensemble
average of the random transverse magnetic field $\vec B_{t}$
squared, $G=6.67\times 10^{-8} \hbox{ dyne cm}^2\hbox{ g}^{-2}$ is
the gravitational constant; the magnetic induction equation
%\[
%\frac{\partial}{\partial t}\big(r^{2}<B_{t}^{2}>\big)
%\]
\begin{equation}\label{basic5}
%\quad\quad\quad
\bigg(\frac{\partial}{\partial t} +u\frac{\partial}{\partial
r}\bigg) \big(r^{2}<B_{t}^{2}>\big)+2r^{2}<B_{t}^{2}> \frac{\partial
u}{\partial r}=0\ ,
\end{equation}
and the conservation equation for `specific entropy'\footnote{The
concept of `specific entropy' is here extended to situations where
$\gamma$, simply regarded as an index parameter, is not necessarily
the ratio of actual gas specific heats. }
\begin{equation}\label{basic6}
\left(\frac\partial{\partial t}+u\frac\partial{\partial
r}\right)\left(\ln\frac p{\rho^\gamma}\right)=0\
\end{equation}
along streamlines. Note that the Poisson equation relating the
mass density and the gravitational potential is consistently
satisfied under the approximation of a quasi-spherical symmetry.
%Here, $\rho$ is the density of the fluid; $u$ is the velocity; $M$
%is the total enclosed mass within radius $r$; $p$ is the thermal
%pressure; $<B_t^2>$ is the mean square of the transverse magnetic
%field; $G$ is the gravitational constant; and $\gamma$ is the
%polytropic index.
This set of MHD equations is the same as that of Wang \& Lou
(2007) and Lou \& Wang (2007), except for equation (\ref{basic6}),
which is the key difference between this analysis and our former
analyses. In other words, the equation of state
$p=\kappa\rho^{\gamma}$ with $\kappa$ being a constant is only one
special case satisfying equation (\ref{basic6}) and is referred to
as the `usual' or `conventional' polytropic equation of state.
Equation of state (\ref{basic6}) has been adopted in various
previous works as noted in Section 1. The assumptions on random
magnetic field distribution (Zel'dovich \& Novikov 1971) and
magnetic induction are the same as those in Yu \& Lou (2005), Yu
et al. (2005), Lou \& Wang (2007) and Wang \& Lou (2007). The work
of Yu \& Lou (2005) studies the free-fall collapse of an
isothermal magnetized gas cloud (Shu 1977; Chiueh \& Chou 1994)
and is relevant to star formation and to formation of hot bubbles
as well as superbubbles in ISM. In contexts of shocks and
`champagne flows' (Tsai \& Hsu 1995; Shu et al. 2002; Bian \& Lou
2005), the MHD model of Yu et al. (2005) extends the scenario and
analysis to magnetized isothermal clouds. The models of Lou \&
Wang (2006, 2007) are designed for rebound shock process in
supernovae without or with a random magnetic field and explore the
origin of intense magnetic fields on compact objects. Given the
above approximations, a proper combination of equations
$(\ref{basic1})-(\ref{basic6})$ leads to an MHD energy
conservation [see equation (7) of Wang \& Lou (2007) and equation
(4.2) of Fan \& Lou (1999)].
%and should not make any difference in here.

By performing the time reversal transformation $t\rightarrow-t$,
$r\rightarrow r$, $u\rightarrow -u$ $\rho\rightarrow\rho$,
$p\rightarrow p$, $M\rightarrow M$ and
$<B^2_t>\rightarrow<B^2_t>$, the above MHD equations are
invariant. Using this time reversal invariant property one can
construct complete global solutions from semi-complete global
solutions (e.g., Lou \& Shen 2004).
%{\bf Should discuss the time reversal property
%of these MHD equations. }{\it to YQL: done.}

\subsection[]{MHD Self-Similarity Transformation}

In order to transform nonlinear MHD partial differential equations
$(\ref{basic1})-(\ref{basic6})$ to a set of nonlinear MHD ordinary
differential equations (ODEs), we introduce the following MHD
self-similarity transformation, namely
\[
r=k^{1/2}t^{n}x\ ,\qquad u=k^{1/2}t^{n-1}v\ ,\qquad
\rho=\frac{\alpha}{4\pi Gt^{2}}\ ,
\]
\begin{equation}\label{transform1}
p=\frac{kt^{2n-4}}{4\pi G}\beta\ ,
M=\frac{k^{3/2}t^{3n-2}m}{(3n-2)G}\ , <B_t^2>=\frac{kt^{2n-4}}{G}w\
.
\end{equation}
Here, $x$ is a dimensionless independent variable, and $v(x)$,
$\alpha(x)$, $\beta(x)$, $m(x)$ and $w(x)$ are functions of $x$
only; $k$ is a dimensional scaling factor to consistently make
$x$, $v$, $\alpha$, $\beta$, $m$ and $w$ dimensionless; $n$ is an
important scaling parameter as noted in Section 1, which
essentially controls the scaling of dimensional physical
quantities with respect to time $t$ and radius $r$. As usual, we
refer to $v(x)$, $\alpha(x)$, $\beta(x)$, $m(x)$ and $w(x)$ as the
reduced radial flow speed, mass density, thermal gas pressure,
enclosed mass, and magnetic energy density (associated with the
completely random transverse magnetic field), respectively.

Under self-similarity transformation (\ref{transform1}), equations
(\ref{basic2}) and (\ref{basic3}) together give the reduced
enclosed mass $m(x)$ as
\begin{equation}\label{transform2}
m=\alpha x^2(nx-v)\ ,
\end{equation}
where $nx-v>0$ is required in order to ensure that the total
enclosed mass remains positive (see Appendix \ref{posdef} for more
details). Also under the same transformation equation
(\ref{transform1}), mass conservation equation (\ref{basic1})
becomes
\begin{equation}\label{transform3}
(nx-v)\alpha'-\alpha v'=-2(x-v)\alpha/x\ ,
\end{equation}
while magnetic induction equation (\ref{basic5}) reads
\begin{equation}\label{transform4}
(nx-v)w'-2wv'=2[v-(2-n)x]w/x\ .
\end{equation}
Equations (\ref{transform3}) and (\ref{transform4}) together lead to
the first integral
\begin{equation}\label{transform5}
w=h\alpha^2x^2\ ,
\end{equation}
where the integration constant $h$ is a dimensionless parameter
representing the strength of random transverse magnetic field
$\vec B_t$. In essence, relation (\ref{transform5}) physically
corresponds to the magnetic frozen-in condition. The above
treatment parallels those of Wang \& Lou (2007) and of Yu \& Lou
(2005), and the results are the same. Combining equations
(\ref{transform2}) and (\ref{transform3}), the `specific entropy'
conservation equation (\ref{basic6}) along streamlines is
equivalent to
\begin{equation}\label{transform6}
\beta=\alpha^\gamma m^q\ ,
\end{equation}
where exponent parameter $q\equiv 2(n+\gamma-2)/(3n-2)$ and hence
$\gamma=2-n+(3n-2)q/2$. In general, equations (\ref{basic2}) and
(\ref{basic6}) imply that $p\rho^{-\gamma}$ (directly related to
the `specific entropy') can be a fairly arbitrary function of the
enclosed mass $M(r,t)$ along streamlines. It is our requirement of
self-similarity solutions that leads to the specific power-law
form of relation (\ref{transform6}). The key point of this
analysis is that $q\neq 0$ or $n+\gamma\neq 2$ is allowed in
general. As noted by Lou \& Cao (2008), here we do not need a
proportional coefficient in equation (\ref{transform6}) for
$\gamma\neq 4/3$. The reason is simply that if we write
\begin{equation}\label{add2}
\beta=\mathcal C\alpha^\gamma m^q\ ,
\end{equation}
where ${\cal C}$ is an arbitrary scaling coefficient, then a
change of parameter $k$ in self-similarity transformation equation
(\ref{transform1}) to $\mathcal C^{1/(1-3q/2)}k$ would make the
coefficient $\mathcal C$ disappear.
%{\bf Do we need a proportional coefficient here in
%general?} {\it to YQL: no. I have added an explanation.}
We shall not consider the special case of $\gamma=4/3$ or $q=2/3$,
in which there exists yet another parameter as an multiplicative
scaling factor between $\beta$ and $\alpha^\gamma m^q$ (see Lou \&
Cao 2008 for a more detailed analysis on the special case of
$\gamma=4/3$ which substantially generalized the earlier research
analysis of Goldreich \& Weber 1980 and Yahil 1983).
%The mass conservation (\ref{basic2}) and the conservation of
%specific entropy (\ref{basic6}) indicate that the specific
%entropy is a function of the enclosed mass along streamlines. This
%general result takes the form of relation (\ref{transform6}) in
%terms of MHD self-similar transformation (\ref{transform1}).
Taking equations (\ref{transform2}), (\ref{transform5}) and
(\ref{transform6}) together, we obtain the following MHD ODE
\[
[\alpha^{2-n+3nq/2}x^{2q}(nx-v)^q]'/\alpha -(nx-v)v'+hx^2\alpha'
\]
\begin{equation}\label{transform7}
\ \qquad=-(n-1)v-2h\alpha x-(nx-v)\alpha/(3n-2)\ .
\end{equation}
Equations (\ref{transform3}) and (\ref{transform7}) together lead to
separate equations for $\alpha'$ and $v'$ respectively in the
compact form of
\begin{equation}\label{transform8}
X(x,\alpha,v)\alpha'=A(x,\alpha,v)\ ,\ \
X(x,\alpha,v)v'=V(x,\alpha,v)\ ,
\end{equation}
where the three functional coefficients $X$, $A$, and $V$ are
defined explicitly below, namely
\[
X(x,\alpha,v)\equiv \bigg(2-n+\frac{3n-2}2q\bigg)
\alpha^{1-n+3nq/2}x^{2q}(nx-v)^q
\]
\[
\qquad\qquad\qquad+h\alpha x^2-(nx-v)^2\ ,
\]
\[
A(x,\alpha,v)\equiv 2\frac{(x-v)}x\alpha
\big[q\alpha^{1-n+3nq/2}x^{2q}(nx-v)^{q-1}
\]
\[
\qquad
+(nx-v)\big]-\alpha\bigg[(n-1)v+\frac{(nx-v)}{(3n-2)}\alpha+2h\alpha
x
\]
\[
\qquad+q\alpha^{1-n+3nq/2}x^{2q-1}(nx-v)^{q-1}(3nx-2v)\bigg]\ ,
\]
\[
V(x,\alpha,v)\equiv
2\frac{(x-v)}x\alpha\bigg[\bigg(2-n+\frac{3n}2q\bigg)
\]
\[
\qquad\quad\times\alpha^{-n+3nq/2}x^{2q}(nx-v)^q+hx^2\bigg]
\]
\[
\qquad\quad-(nx-v)\bigg[(n-1)v +\frac{(nx-v)}{(3n-2)}\alpha+2h\alpha
x
\]
\begin{equation}\label{transform9}
\qquad+q\alpha^{1-n+3nq/2}x^{2q-1}(nx-v)^{q-1}(3nx-2v)\bigg]\ .
\end{equation}
Equations (\ref{transform8}) and (\ref{transform9}) are coupled
nonlinear MHD ODEs and with specified `boundary' and `initial'
conditions (i.e., appropriate asymptotic solutions at large and
small $x$), they can be integrated numerically by the standard
fourth-order Runge-Kutta method (e.g., Press et al. 1986); one
needs to pay special attention on the {\it magnetosonic singular
surface} where $X(x,\alpha, v)=0$ in both ODEs of
(\ref{transform8}). The solutions of these two coupled nonlinear
MHD ODEs cannot smoothly cross the singular surface unless they
cross it with a MHD shock, or they satisfy the critical condition
on the so-called magnetosonic critical curve where both the
denominator $X(x,\alpha,v)$ and the numerators $A(x,\alpha,v)$ and
$V(x,\alpha,v)$ of these two equations in (\ref{transform8})
vanish simultaneously. The magnetosonic critical curve can be
determined numerically by the specific procedure described in
Appendix \ref{cridet}, and the magnetosonic critical condition is
derived and solved numerically in Appendix \ref{eigsol}. As
necessary checks, our analysis here reduces to earlier known
results (e.g., Shu 1977; Suto \& Silk 1988; Lou \& Shen 2004; Yu
\& Lou 2005; Bian \& Lou 2005; Lou \& Wang 2006; Wang \& Lou 2007)
if we set the relevant parameters to appropriate values (e.g.,
$q=0$, $\gamma=n=1$, and $h=0$, and so forth).

\subsection[]{MHD Shock Conditions in Self-Similarity Form}

An MHD shock is characterized by discontinuities in thermal gas
pressure, mass density, temperature, tangential magnetic field and
radial flow velocity, caused by steepening of nonlinear MHD flow
evolution or by various explosion and jet processes. In our
self-similar and quasi-spherical symmetric formulation for a
general polytropic gas, a self-similar MHD shock can be readily
constructed mathematically, which represents a possible evolution
of an MHD shock in such a magnetized flow system (see e.g.,
Chevalier 1982; Tsai \& Hsu 1995; Shu et al. 2002; Shen \& Lou
2004; Bian \& Lou 2004; Yu et al. 2006; Lou \& Wang 2007 for
earlier results on self-similar shocks either with or without a
completely random magnetic field).
%In fact, alike the fluid system itself, shocks may also
%evolve into a self-similar shock together with the whole system.

To characterize an MHD shock specifically, we need to impose several
conservation laws across the two sides of a shock front in the
comoving framework of reference. Denoting $u$ as the flow velocity,
$u_s$ as the shock velocity, $\rho$ as the gas mass density, $p$ as
the thermal gas pressure, and $<B_t^2>$ as the mean square of the
random transverse magnetic field, respectively, we have in the shock
framework of reference the mass conservation
\begin{equation}\label{shock1}
\big[\rho(u_s-u)\big]_1^2=0\ ,
\end{equation}
the radial momentum conservation
\begin{equation}\label{shock2}
\bigg[p+\rho(u_s-u)^2+\frac{<B_t^2>}{8\pi}\bigg]_1^2=0\ ,
\end{equation}
the MHD energy conservation equation
\begin{equation}\label{shock3}
\bigg[\frac{\rho(u_s-u)^3}{2}+\frac{\gamma p(u_s-u)}
{(\gamma-1)}+\frac{<B_t^2>}{4\pi}(u_s-u)\bigg]_1^2=0\ ,
\end{equation}
and the magnetic induction equation
\begin{equation}\label{shock4}
\big[(u_s-u)^2<B_t^2>\big]_1^2=0\ .
\end{equation}
%where $[\ \ ]_1^2$ represents the difference of physical
%quantities across a shock in the comoving framework of reference.
This set of jump conditions for an MHD shock is the same as that
adopted in Lou \& Wang (2007). We here use a pair of square
brackets outside each expression enclosed to denote the difference
between the upstream (marked by subscript `1') and downstream
(marked by subscript `2') quantities, as has been done
conventionally for shock analyses (Landau \& Lifshitz 1959;
Zel'dovich \& Raizer 1966, 1967).

There are two parallel sets of MHD self-similar transformation in
the upstream and downstream domains respectively, because the
parameter $k$ in self-similarity transformation equation
(\ref{transform1}) is related to the sound speed and thus can be
different in the two flow regions across a shock (see e.g., Lou \&
Wang 2006 and Lou \& Gao 2006). We shall set $k_2=\lambda^2k_1$
with the scaling ratio $\lambda$ representing this difference in
the upstream $k_1$ and the downstream $k_2$. The following
relations are necessary for consistency
\begin{equation}\label{shock5}
\qquad h_1=h_2\ ,\qquad\qquad\qquad x_1=\lambda x_2\ ,
\end{equation}
and we shall simply use $h$ instead of $h_1$ and $h_2$ from now
on. Using the two relations in (\ref{shock5}), MHD shock jump
conditions (\ref{shock1}) to (\ref{shock4}) can be readily cast
into the following self-similarity form
\begin{equation}\label{shock6}
\alpha_1(nx_1-v_1)=\lambda\alpha_2(nx_2-v_2)\ ,
\end{equation}
\[
\alpha_1^{2-n+3nq/2}x_1^{2q}(nx_1-v_1)^{q}+\alpha_1(nx_1-v_1)^2
\]
\[
\qquad+\frac{h\alpha_1^2x_1^2}{2}
=\lambda^2\bigg[\alpha_2^{2-n+3nq/2}x_2^{2q}(nx_2-v_2)^q
\]
\begin{equation}\label{shock7}
\qquad\qquad\qquad\qquad\qquad
+\alpha_2(nx_2-v_2)^2+\frac{h\alpha_2^2x_2^2}{2}\bigg]\ ,
\end{equation}
\[
(nx_1-v_1)^2+\frac{2\gamma}{(\gamma-1)}
\alpha_1^{1-n+3nq/2}x_1^{2q}(nx_1-v_1)^{q}
\]
\[
\qquad\qquad\quad+2h\alpha_1x_1^2=\lambda^2\bigg[(nx_2-v_2)^2
\]
\begin{equation}\label{shock8}
\quad+\frac{2\gamma}{(\gamma-1)}
\alpha_2^{1-n+3nq/2}x_2^{2q}(nx_2-v_2)^q+2h\alpha_2x_2^2\bigg]\ .
\end{equation}
These three self-similar MHD shock equations (\ref{shock6})$-$
(\ref{shock8}) can be explicitly solved and relevant details of
derivation can be found in Appendix \ref{shocksolve}.

\section[]{Asymptotic MHD Solutions}
%\section[]{Model Analysis}\label{math}

We connect relevant asymptotic MHD solutions by numerical
integration to construct semi-complete global solutions for
self-similar MHD flows with or without shocks. Asymptotic
analytical MHD solutions are extremely valuable and illustrate the
balance and dominance of various forces such as gravity, pressure
force and Lorentz force. All MHD solutions obtained here are more
general with $q\neq 0$ (i.e., $n+\gamma\neq 2$) and are consistent
with previous results published in the literature (Shu 1977; Suto
\& Silk 1988; Lou \& Shen 2004; Yu \& Lou 2005; Wang \& Lou 2007;
Lou \& Wang 2006, 2007; Lou \& Li 2007 in preparation; Lou \& Cao
2008).

\subsection[]{MHD Solutions of Pressure Dominance}

When the thermal pressure becomes significant in an asymptotic MHD
solution, it is generally expected that the pressure term will enter
the asymptotic ODEs governing this solution. Therefore, parameter
$q$ will appear in the leading order of the asymptotic MHD solution.

\subsubsection[]{Asymptotic MHD Solutions of
Finite Density and Velocity at Large $x$}

For asymptotic MHD solutions at large $x$, the gravitational force,
magnetic force and thermal pressure gradient force are all in the
same order of magnitude, and $v(x)$ and $\alpha(x)$ become finite or
approach zero in the limit of $x\rightarrow +\infty$. In this case
of large $x$, we have to leading order the following coupled
nonlinear MHD ODEs
\[
\alpha'=-\frac{2\alpha}{nx}\ ,
\]
\[
v'=\frac{(n-1)v}{nx}+\bigg[\frac{1}{(3n-2)}
+\frac{2h(n-1)}{n^2}\bigg]\alpha
\]
\begin{equation}\label{asym7}
\qquad\qquad-2(2-n)n^{q-2}\alpha^{1-n+3nq/2}x^{3q-2}\ ,
\end{equation}
and the corresponding asymptotic MHD solution is
\begin{equation}\label{asym8}
\alpha=Ax^{-2/n}+\cdots\ ,
\end{equation}
\[
v=Bx^{1-1/n}+\bigg\{-\bigg[\frac{n}{(3n-2)}
+\frac{2h(n-1)}{n}\bigg]A
\]
\begin{equation}\label{asym9}
\qquad\qquad +2(2-n)n^{q-1}A^{1-n+3nq/2}\bigg\}x^{1-2/n}+\cdots\ ,
\end{equation}
where $A$ and $B$ are two constants of integration. The radial
profile of magnetic field can be readily derived from equation
(\ref{transform5}). We find that inequality $n>2/3$ (i.e., a
positive enclosed mass $M$) is sufficient to warrant the validity of
this asymptotic MHD solution. For $2/3<n<1$, the two coefficients
$A$ and $B$ can be fairly arbitrary.
%{\bf How about $n<2/3$? Please discuss or comment.}
The special isothermal case (Shu 1977) corresponds to $B=0$, $q=0$,
$n=1$, $\gamma=1$, and $h=0$;
% was first derived and utilized by Shu (1977),
and the $B\neq 0$, $q=0$, $n=1$, $\gamma=1$, $h=0$ version was
implicit in Hunter (1977).
%{\bf Any other relevant comments and
%comparison?}{\it to YQL: no.}
The constant coefficient $B$ is a free parameter in this solution,
because $v=Bx^{1-1/n}$ itself would satisfy $\partial u/\partial t
+u\partial u/\partial r=0$, as if no force is active. For $n=1$, the
flow approaches a constant radial velocity at the outer portion
(Whitworth \& Summers 1985; Lou \& Shen 2004). This asymptotic
solution has been applied to various collapse problems (e.g. Shu
1977; Yahil 1983; Lou \& Shen 2004; Bian \& Lou 2005; Lou \& Wang
2006; Lou \& Gao 2006), because of the common feature of inflow,
outflow, contraction and/or collapse of dynamic evolutions. Note
that $n=1$ case here differs from that of Suto \& Silk (1988)
because we require specific entropy conservation along streamlines.
For $1<n<2$, we should set $B=0$ to avoid the flow speed divergence
of the first term in solution (\ref{asym9}) at large $x$. With $n=2$
and $B=0$, we have a constant asymptotic inflow speed again by
solution (\ref{asym9}). For $n>2$, we must require both leading
order terms in flow speed solution (\ref{asym9}) to vanish to avoid
velocity divergence; nevertheless for coefficient $A>0$, the second
term in flow speed solution (\ref{asym9}) does not vanish; this
second term would vanish for a trivial solution of $A=0$. Unless
other constraints appear, we must require $2/3<n\leq 2$,
corresponding to a range of mass density scaling $\rho\propto
r^{-3}$ to $\rho\propto r^{-1}$. For a comparison, Cheng (1978)
indicated the more obvious limit of $\rho\propto r^{-3}$ for
$n\rightarrow 2/3$, while our analysis here also gives another limit
of $\rho\propto r^{-1}$ for $n\rightarrow 2$ due to the second term
in asymptotic solution (\ref{asym9}) for the flow speed. At large
$x$, the temperature scales as $x^{2(n-1)/n}$. For $n\leq 1$, the
temperature remains finite, while for $n>1$, the temperature
diverges as $x\rightarrow\infty$ and one needs to invoke a finite
size of an astrophysical system in order to make use of this
solution.

%{\bf Weigang: please check my added analysis and comments
%above.}{\it to YQL: OK.}

\subsubsection[]{Magnetostatic Solution for a Polytropic Sphere}

Setting $v=0$ for all $x$ in equation (\ref{transform8}), one
immediately obtains an exact magnetostatic solution\footnote{In
equation (11) of Lou \& Wang (2006), there are two typos; in the
coefficients of both $\alpha(x)$ and $m(x)$ there, the exponent
should be $-1/n$ instead of $1/n$. }
\[
\alpha=\bigg[\frac{n^2-2(1-n)(3n-2)h}{2(2-n)(3n-2)}
n^{-q}\bigg]^{-1/(n-3nq/2)}x^{-2/n}
\]
\begin{equation}\label{qstatic1}
\ \ \ \equiv A_0x^{-2/n}\ .
\end{equation}
In the limit of $v\rightarrow 0$, we may still regard equation
(\ref{transform5}) to be valid from the MHD evolutionary
perspective; it is then straightforward to derive the
corresponding radial profile of the random magnetic field. This
magnetostatic solution for a singular polytropic sphere (SPS) is
globally valid for parameters that physically allow expression
(\ref{qstatic1}) except for the central divergence as
$x\rightarrow 0^+$. The key difference here is that $q\neq 0$ or
$n+\gamma\neq 2$ is allowed in general in SPS solution
(\ref{qstatic1}). The quasi-magnetostatic solution described
immediately below is constructed based on SPS solution
(\ref{qstatic1}). Of course, all forces present are significant in
this solution.

\subsubsection[]{Quasi-Magnetostatic Solutions at Small $x$}

For quasi-magnetostatic asymptotic solutions (see Lou \& Wang
2006, 2007 and Wang \& Lou 2007 for details of deriving the
hydrodynamic and MHD counterparts with $q=0$; see also Lou, Jiang
\& Jin 2008 and Jiang \& Lou [2008, in preparation] for this
solution in two gravity-coupled fluids) at small $x$ values, we
first substitute
\begin{equation}\label{qstatic2}
\alpha=A_0x^{-2/n}+Nx^{K-1-2/n}\ ,\ \qquad v=Lx^K\ ,
\end{equation}
into coupled nonlinear MHD ODEs (\ref{transform8}) and
(\ref{transform9}), where $A_0$ is explicitly defined by equation
(\ref{qstatic1}), and then obtain two nonlinear algebraic equations
for the three relevant parameters $N$, $K$ and $L$, namely
\begin{equation}\label{qstatic3}
n(K-1)N=\big(K+2-2/n\big)A_0L\ ,
\end{equation}
\[
\bigg[\frac{QK}n+\frac{1}{(3n-2)}
+\bigg(\frac1n-\frac4{n^2}\bigg)Q\bigg]A_0L
\]
\[
\qquad =\bigg\{\bigg[\frac{n^2}{2(3n-2)} +nh+\frac{3n}2Q\bigg]K
\]
\begin{equation}\label{qstatic4}
\quad\qquad+\frac{n^2}{2(3n-2)}-(2-n)h
+\bigg(\frac{3n}2-6\bigg)Q\bigg\}N\ .
\end{equation}
Here we have introduced a handy parameter
\begin{equation}\label{qstatic5}
Q\equiv q\bigg[\frac{n^2}{2(2-n)(3n-2)} -\frac{(1-n)}{(2-n)}h\bigg]
\end{equation}
for simplicity and for the convenience of mathematical derivation.
Equations (\ref{qstatic3}) and (\ref{qstatic4}) together give rise
to a quadratic equation for $K$ in terms of three parameters $n$,
$Q$ and $h$, namely
\[
\bigg[\frac{n^2}{2(3n-2)}+nh+\frac{(3n-2)}{2}Q\bigg]
\bigg[K^2+\frac{(3n-4)}nK\bigg]
\]
\begin{equation}\label{qstatic6}
\ +\frac{2(2-n)(1-n)}{n}h +\frac{n^2+(3n-2)^2(1-4/n)Q}{(3n-2)}=0\ .
\end{equation}
Once proper roots of $K$ are known, parameters $N$ and $L$ are
proportional to each other by equation (\ref{qstatic3}) and only
one of them is free to choose. The existence of the exact
magnetostatic SPS solution (\ref{qstatic1}) as well as the
requirement of $\Re(K)>1$ constrain the parameter regime of this
quasi-magnetostatic solution. In reference to equations (6) and
(7) of Lou \& Wang (2007) corresponding to $Q=0$ here, we have
explicitly discussed the parameter regime where two real roots
$K>1$ exist. We therefore expect that at least for a sufficiently
small $Q\neq 0$, it would be still possible to have two real roots
of $K>1$ given by quadratic equation (\ref{qstatic6}). Of course,
it is also possible to have one root $K>1$ and the other root
$K<1$. For example, for $\gamma=1.4$, $n=1.4$ and $h=0$, we have a
positive $K=1.056$ and a negative $K=-1.199$; with other
parameters the same but $h=1$, the two $K$ roots are $K=1.0373$
and $-1.180$, respectively.

%{\bf How about the other $K$?}{\it to YQL: the other $K$ equals
%-1.199} {\bf Please provide a few examples of $h\neq 0$ if
%possible.} {\it to YQL: for $h=1$, $K=1.0373, -1.180$ (other
%parameters are the same)}
%
%{\bf Have you explored possible $K$ solutions
%as concrete examples?}{\it to YQL: done.}
%{\bf Any comment on Larson-Penston type
%solutions at small $x$? No longer exist! }

\subsubsection[]{MHD Thermal Fall Solutions at Small $x$}

Lou \& Li (2007 in preparation) report a possible asymptotic thermal
fall solution for small $x$ without magnetic field; for this
solution, the thermal pressure gradient force almost balances the
gravitational force, yet the mass density and radial infall speed
still approach infinity at small $x$. In the case of asymptotic MHD
thermal fall solutions, magnetic force is much smaller than both the
pressure gradient force and the gravity at small $x$. The leading
nonlinear ODEs for small $x$ read
\begin{equation}\label{thermalfall1}
vv'+{\beta'}/{\alpha}-{v\alpha}/{(3n-2)}=0\ ,
\end{equation}
\begin{equation}\label{thermalfall2}
(v\alpha)'+2v\alpha/x^2=0\ ,
\end{equation}
where the effect of magnetic field does not appear explicitly (i.e.,
the absence of $h$ parameter) in the regime of small $x$. This pair
of coupled nonlinear ODEs (\ref{thermalfall1}) and
(\ref{thermalfall2}) at small $x$ are the same as those in Lou \& Li
(2007 in preparation), and the corresponding asymptotic solution is
therefore the same to the leading order, namely
\begin{equation}\label{thermalfall3}
\alpha=\bigg[\frac{\gamma(3n-2)}{(\gamma-1)}\bigg]^{-1/(\gamma-1)}
m(0)^{-{(q-1)}/{(\gamma-1)}}x^{-1/{(\gamma-1)}}\ ,
\end{equation}
\[
v=-\bigg[\frac{\gamma(3n-2)}{(\gamma-1)}\bigg]^{1/(\gamma-1)}
m(0)^{(q+\gamma-2)/(\gamma-1)}
\]
\begin{equation}\label{thermalfall4}
\qquad\qquad\qquad\qquad\qquad\qquad \qquad \times
x^{-(2\gamma-3)/(\gamma-1)}\ .
\end{equation}
Here $m(0)$ is the value of the total reduced enclosed mass $m(x)$
as $x$ approaches 0, representing an increasing point mass at the
very centre. This asymptotic solution at small $x$ is valid for
$3/2<\gamma<5/3$. The corresponding radial profile of the random
magnetic field can be readily inferred from equation
(\ref{transform5}).

\subsubsection[]{Solutions of MHD Thermal Expansion at Large $x$}

The asymptotic MHD thermal fall solutions  (\ref{thermalfall3})
and (\ref{thermalfall4}) at small $x$ in the preceding subsection
{\it 3.1.4} can be integrated numerically and connected with the
asymptotic MHD thermal expansion solution at large $x$ (see Figure
\ref{thermalfall}). For this latter solution at large $x$, the
reduced radial flow velocity $v(x)$ becomes linear in $x$ while
the mass density approaches zero for $x\rightarrow\infty$. The
force that is dominant here is the thermal pressure gradient,
driving the gas into expansion. In this case of $v\sim cx$ with
$c$ being a constant coefficient, we obtain from coupled nonlinear
ODEs (\ref{transform3}) and (\ref{transform7}) at large $x$
\begin{equation}\label{thermalex1}
(n-c)x\alpha'-c\alpha=-2(1-c)\alpha\ ,
\end{equation}
\begin{equation}\label{thermalex2}
(n-1)cx-(n-c)cx+{\beta'}/{\alpha}=0\ .
\end{equation}
By further assuming the power-law form of $\alpha\sim Ex^P$ at large
$x$ with $E$ and $P$ being two parameters and using equations
(\ref{transform2}) and (\ref{transform6}), we obtain from equations
(\ref{thermalex1}) and (\ref{thermalex2})
\begin{equation}\label{thermalex3}
P=-{(3q-2)}/{(1-n+3nq/2)}\ ,
\end{equation}
\begin{equation}\label{thermalex4}
E^{1-n+3nq/2}(n-c)^q(2+P)=c(1-c)\ ,
\end{equation}
and
\begin{equation}\label{thermalex5}
P={(3c-2)}/{(n-c)}\ .
\end{equation}
Parameters $P$, $c$ and $E$ are then readily determined by three
algebraic equations (\ref{thermalex3})$-$(\ref{thermalex5}) once
the values of $q$ and $n$ parameters are specified. One can easily
fix $P$ from equation (\ref{thermalex3}) first, then calculate $c$
from equation (\ref{thermalex5}) and finally obtain $E$ by
equation (\ref{thermalex4}). This solution for MHD thermal
expansion is valid when $\gamma>4/3$ (i.e., $q>2/3$), because we
require $P<0$ for a converging $\alpha(x)$ at large $x$. Again,
the corresponding radial profile of magnetic field can be readily
inferred from equation (\ref{transform5}).

Asymptotic solution here should be compared with asymptotic
solutions (\ref{asym8}) and (\ref{asym9}) at large $x$ when $n<1$.
%{\bf Any obvious reason for this inequality?}{\it to YQL: done.}
%
%{\bf How about the possibility of an exact
%global solution? No longer exist! }

\subsection[]{MHD Solutions with Weak Thermal Pressure }

For MHD solutions involving weak thermal pressure, parameter $q$
should not enter the leading order terms of the asymptotic solution.

\subsubsection[]{MHD Free-Fall Solutions at Small $x$ }

For asymptotic central free-fall solutions at small $x$, first
found by Shu (1977) under the isothermal approximation, the
gravity force is virtually the only force in action, and the
radial velocity and mass density profiles both diverge in the
limit of $x\rightarrow 0^+$. The leading order ODEs at small $x$
then read
\[
\alpha'=-\frac{2\alpha}{x}-\frac{\alpha^{2}}{(3n-2)v}\ ,
\]
\begin{equation}\label{asym1}
v'=\frac{\alpha}{(3n-2)}\ ,
\end{equation}
and the corresponding asymptotic solution appears as
\begin{equation}\label{asym2}
\alpha(x)=\bigg[\frac{(3n-2)m(0)}{2x^{3}}\bigg]^{1/2}\ ,
\end{equation}
\begin{equation}\label{asym3}
v(x)=-\bigg[\frac{2m(0)}{(3n-2)x}\bigg]^{1/2}\ ,
\end{equation}
with an integration constant $m(0)$, representing an increasing
point mass at the very centre. The special isothermal case of
$n=1$ and $\gamma=1$ was first studied by Shu (1977) in the
context of star formation (Shu et al. 1987). It is particularly
interesting to note that it is now possible to have $\gamma=1$ and
$n>2/3$ case, still corresponding to a non-isothermal gas flow
(i.e., $q\neq 0$ and a variable sound speed). On the other hand,
it is also possible to have $n=1$ and $\gamma>1$ case with $q\neq
0$. With various combinations of sensible parameters, we can now
readily construct such kind of solutions numerically.
%to specifically illustrate this point. {\bf Anything interesting
%or strange for the case of $\gamma=1$ and $n\neq 1$?}
%{\bf wwg has constructed numerical solutions. }
In terms of modeling the MHD processes of star or core formation
in magnetized clouds (Zhou et al. 1992; Shen \& Lou 2004; Fatuzzo
et al. 2004; Lou \& Gao 2006), our general polytropic MHD model
framework is more versatile including the possible role of a
completely random transverse magnetic field. In particular, we can
model various molecular spectral line profiles in star forming
regions; we shall pursue this exploration in separate papers.

To the leading order, this asymptotic MHD free-fall solution does
not involve polytropic index $\gamma$ and the corresponding
profile of magnetic field can be readily determined by equation
(\ref{transform5}). In this case, magnetic field does not play a
dynamically important role but may reveal diagnostic features if
shock accelerated relativistic electrons are present. We now
consider the parameter regime that allows for this asymptotic MHD
free-fall solution at small $x$. Substituting equations
(\ref{asym2}) and (\ref{asym3}) into coupled nonlinear MHD ODEs
(\ref{transform8}) and (\ref{transform9}) and requiring the
emergence of equation (\ref{asym1}) for small $x$ values with
consistent orders of magnitudes for the higher order terms,
%{\bf need to understand this better}
we obtain two inequalities
\begin{equation}\label{asym10}
n>2/3\qquad\qquad\mbox{and}\qquad\qquad\gamma<5/3\ ;
\end{equation}
these requirements appear to be the same as those of Suto \& Silk
(1988), but the polytropic equation of state adopted is different
between theirs and ours, that is, it is no longer necessary to
impose the condition $n+\gamma=2$ or $q=0$ in our more general MHD
polytropic model formulation.

\subsubsection[]{Strong-Field Asymptotic MHD Solutions in the\\ \qquad\ Small-$x$ Regime}

\begin{figure}
\includegraphics[width=3.3in,bb=100 160 500 630]{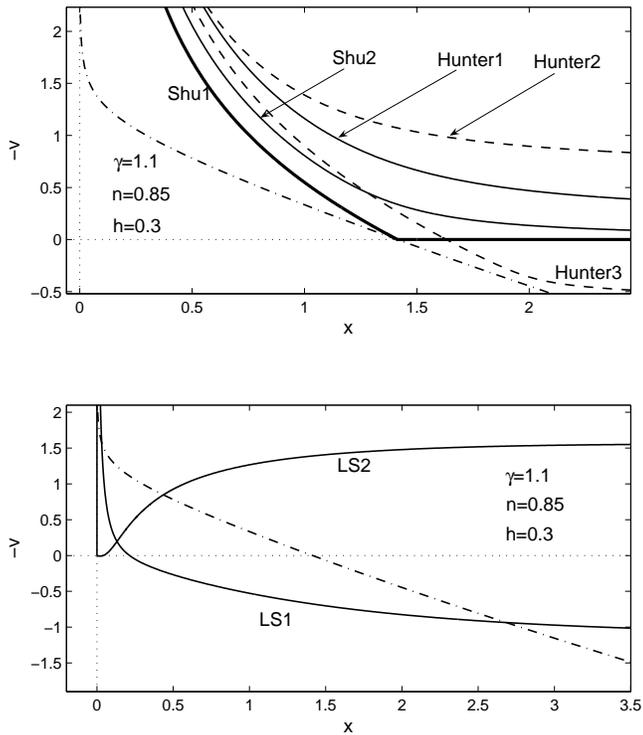}
\caption{
%{\bf It is also possible to put parameters
%in the figure.}{\it to YQL: done.}
Semi-complete MHD solutions with inner free-fall asymptotic
solutions and without MHD shocks. All these solutions involve a
completely random magnetic field and a general polytropic gas. The
relevant parameters are $\gamma=1.1,\ n=0.85,\ q=-0.182$, and
$h=0.3$. The two perpendicular dotted straight lines are abscissa
and ordinate axes, respectively. The dash-dotted curves are the
magnetosonic critical curves. In the upper panel, the MHD solution
labelled with `Shu1' is a polytropic MHD counterpart of the
isothermal expansion-wave collapse solution (EWCS; Shu 1977), and is
constructed by integrating from large $x$ asymptotic solutions
(\ref{asym8}) and (\ref{asym9}) with parameters $A$ slightly larger
than $A_0=1.842$ and $B=0$; the MHD solution labelled with `Shu2' is
constructed in a similar manner for $A=2$ and $B=0$ in asymptotic
solutions (\ref{asym8}) and (\ref{asym9}); the MHD solution labelled
with `Hunter1' is constructed with two parameters $A=2$ and
$B=-0.3$; the `Hunter2' solution is constructed with $A=1.6<A_0$ and
$B=-1$; and the `Hunter3' solution with $A=3$ and $B=1$. We refer to
these MHD counterparts as Hunter type solutions, because Hunter
(1977) implicitly constructed such isothermal solutions without
magnetic field in the complete space.
%{\bf Why do you call it `Hunter'?}{\it to YQL:
In the lower panel, the MHD solution labelled with `LS1' is
constructed by integrating from $(x,v,\alpha)=(0.03092,\ -1.581,\
56.94)$ on the magnetosonic critical curve inwards to small $x$ and
outwards to $x_F=0.3$, and by integrating from
$(x,v,\alpha)=(2.683,\ 0.9359,\ 0.4795)$ on the magnetosonic
critical curve inwards to the same $x_F$ and outwards to large $x$;
this solution goes smoothly across the magnetosonic critical curve
twice. The other MHD solution labelled with `LS2' is constructed by
integrating from $(x,v,\alpha)=(3.631\times 10^{-6},\ -4.944,\
1.497\times10^6)$ on the magnetosonic critical curve inwards to
small $x$ and outwards to $x_F=0.3$, and by integrating from
$(x,v,\alpha)=(0.4357,\ -0.8466,\ 2.8839)$ on the magnetosonic
critical curve inwards to the same $x_F$ and outwards to large $x$
(Lou \& Shen 2004).
% for solution matching procedure].
The values of $m(0)$ for the `Shu1', `Shu2',
`Hunter1',`Hunter2',`Hunter3', `LS1' and `LS2' MHD solutions are
$1.286, 1.565, 1.777, 1.767, 1.864, 0.08699$ and
$9.763\times10^{-5}$, respectively. The `LS1' MHD solution has its
large $x$ asymptotic solution with parameters $A=4.241$ and
$B=1.843$ in asymptotic solution (\ref{asym8}) and (\ref{asym9});
and the `LS2' MHD solution has its large $x$ asymptotic solution
with $A=0.4973$ and $B=-2.431$ in asymptotic solution (\ref{asym8})
and (\ref{asym9}). As examples of illustration, the `LS1' and `LS2'
MHD solutions cross the magnetosonic critical curve twice smoothly.
The `LS2' MHD solution has self-similar oscillations with two nodes
of $v=0$ at very small $x$.} \label{freefallnoshock}
\end{figure}

The strong-field asymptotic MHD solutions in the small-$x$ regime
(see Wang \& Lou 2007 for the $q=0$ counterpart of this asymptotic
solution in a conventional polytropic gas) are central or collapse
solutions with the magnetic Lorentz force against the self-gravity
and with their strengths being comparable in magnitudes; both
forces overwhelm the thermal pressure gradient force. For such
strong-field asymptotic solutions, the velocity profile remains
finite but the mass density profile becomes divergent as
$x\rightarrow 0^{+}$. More specifically in the regime of
$x\rightarrow 0^{+}$, we obtain leading-order terms for the three
coefficients in equation (\ref{transform9}), namely
\[
X(x,\alpha, v)\sim h\alpha x^2\ ,
\]
\[
A(x,\alpha,v)\sim -\frac{(nx-v)}{(3n-2)}\alpha^2-2h\alpha^2x\ ,
\]
\begin{equation}\label{asym4}
V(x,\alpha,v)\sim2h\alpha x^2(1-n)+\frac{(nx-v)^2}{(3n-2)}\alpha\ ,
\end{equation}
such that the two coupled nonlinear MHD ODEs in (\ref{transform8})
can be simplified to
\begin{equation}\label{asym5}
v(x)=\bigg\lbrace n-\frac{(3n-2)}{2} \left[h\pm
(h^2-4h)^{1/2}\right]\bigg\rbrace x\ ,
\end{equation}
and
\begin{equation}\label{asym6}
\alpha(x)=D_0x^{-5/2\mp\sqrt{h^2-4h}/(2h)}\
\end{equation}
to the leading order of small $x$, where $D_0$ is an integration
constant; the corresponding magnetic field strength profile is
simply characterized by equation (\ref{transform5}). For
strong-field asymptotic MHD solutions (\ref{asym5}) and
(\ref{asym6}) to be physically valid, it is necessary to require
that $h>4$, corresponding to a sufficiently strong magnetic field
regime and hence the name of this type of asymptotic MHD
solutions. In a wide range of astrophysical systems, the typical
value of $h$ ranges from $\sim 10^{-3}$ to $\sim 10^5$. The
strong-field situations may happen for magnetospheric plasmas
surrounding magnetic white dwarfs, radio pulsars, anomalous X-ray
pulsars (AXPs), magnetars and so forth. Typical magnetic field
strengths are $\sim 10^7-10^9$G for magnetic white dwarfs, $\sim
10^{9}-10^{10}$G for millisecond radio pulsars, $\sim
10^{11}-10^{12}$G for radio pulsars, $\sim 10^{12}-10^{13}$G for
AXPs, and $\sim 10^{13}-10^{15}$G for magnetars.

To the leading order, the polytropic index $\gamma$ does not
appear in the strong-field asymptotic solution, because the
thermal pressure force is much weaker than both gravity and
magnetic force. By requiring
$\beta=m^q\alpha^{\gamma}\ll\alpha^2x^2$ in the derivation, the
parameter regime for these strong-field asymptotic MHD solutions
at small $x$ to be valid is
\begin{equation}\label{asym11}
\bigg[-\frac{5}{2}\mp\frac{(h^2-4h)^{1/2}}{2h}\bigg]
\bigg(-n+\frac{3nq}{2}\bigg)+3q-2>0\ .
\end{equation}
By setting $q=0$ and thus $n+\gamma=2$, inequality (\ref{asym11})
reduces to inequality (69) of Wang \& Lou (2007) for a
conventional polytropic gas permeated with a completely random
transverse magnetic field.

Numerical examples of using strong-field asymptotic solutions
(\ref{asym5}) and (\ref{asym6}) with the upper signs are shown in
Figure \ref{magnetoaccretion}. These solutions (labelled by `WL1'
and `WL2') pass through the magnetosonic curve smoothly and join the
asymptotic solutions (\ref{asym8}) and (\ref{asym9}) at large $x$.
One can also construct MHD shock solutions in this scheme
numerically.

%{\bf Please check the origin of inequality (\ref{asym11}).} {\bf
%Please provide more detailed analysis of this strong field
%solution for $q\neq 0$.}{\it to YQL: the origin of inequality
%(\ref{asym11}) should be $\beta<<\alpha^2x^2$}

We note empirically that numerical integrations from very small
$x$ outwards using strong-field asymptotic solutions (\ref{asym5})
and (\ref{asym6}) have a tendency to be unstable, especially for
the lower-sign solution. On the other hand, we are unable to
obtain the lower-sign solutions in (\ref{asym5}) and (\ref{asym6})
by numerically integrating from the magnetosonic critical curve to
small $x$ so far. This problem might be related to properties of
perturbations in a self-similar flow (Lou \& Bai 2008 in
preparation).

%{\bf Again, what is the implication of two possible strong field
%asymptotic MHD solutions?}{\it to YQL: the strong field asymptotic
%MHD solutions are unstable in integration directions. Even with an
%exact backward (using results of integrating from large x to small
%x as small x initial values) integration, it is unstable. If we
%use a set of parameter C and D to integrate outward from small x,
%it is extremely unstable.}
%
%{\bf Please clarify this situation. } {\it to YQL: I find that if
%we choose the exact backward integration mentioned in the above
%italic words so that the small x initial value is not too small,
%the backward integration is still quite stable, thus the
%"instability" is a numerical one, not a systematic one. But if we
%use the asymptotic forms as small x initial values, the results
%are systematically different: the obtained solution is not the
%corresponding asymptotic solution, it does not follow the
%asymptotic form immediately as expected, but departs from this
%form very quickly. Thus it is incorrect to integrate from small x
%to large x. The solution with the other root is also the same in
%this situation.}
%
%{\bf Need a further discussion with wwg on this.}

Lou \& Wang (2007) proposed a class of self-similar MHD rebound
shock models for supernovae based on quasi-static MHD shock
solutions with $q=0$ and focussed on the origin(s) of intense
magnetic fields on remnant compact objects such as radio pulsars
(magnetic field strengths of $\sim 10^{11}-10^{12}$G) and magnetic
white dwarfs (surface magnetic field strengths of $\sim
10^{7}-10^{9}$G). It can be possible that MHD rebound shock models
based on our strong-field solutions (see Figure
\ref{magnetoaccretion}) with $q\neq 0$ are relevant to strong
magnetic fields of $\sim 10^{13}-10^{15}$G observationally
estimated for magnetars and anomalous X-ray pulsars (AXPs). This
would correspond to strong surface magnetic field strengths higher
than several thousand gauss on progenitor stars (e.g., magnetic Ap
stars). We shall discuss this interesting problem in separate
papers.

\section[]{Global Semi-Complete
MHD Solutions\\ \quad and Astrophysical Applications}\label{appl}

\subsection[]{MHD Expansion-Wave Collapse Solutions}
%
%\section{Global Similarity Solutions}\label{numerical}

Shu (1977) constructed the expansion-wave collapse solution (EWCS)
for a self-similar {\it isothermal} gas flow and later developed
the so-called inside-out collapse scenario for protostar formation
(see Shu, Adams \& Lizano 1987 and extensive references therein).
Cheng (1978) presented the polytropic generalization of EWCS.
Chiueh \& Chou (1994) studied the MHD generalization of an
isothermal EWCS. In reference to the work of Yu \& Lou (2005) in
this context, we can now construct the general polytropic MHD EWCS
counterpart here and show a specific example labelled with `Shu1'
in Figure \ref{freefallnoshock}. In our general polytropic MHD
EWCS, the outer portion is the outer part of a magnetized SPS (no
singularity is actually involved because $x=0$ is excluded), while
the inner portion approaches an MHD asymptotic free-fall solution
with the random magnetic field being advected radially inward.
From the perspective of theoretical model development, we can now
test thermodynamic properties of magnetized clouds by comparing
with observational inferences. Using the isothermal version of
this solution, Shu (1977) outlined a physical scenario for
protostar formation, which can now be further extended on the
basis of our generalized polytropic MHD model solution. At the
beginning, the magnetized SPS is static; then at $t=0$, a
disturbance may take place and a magnetized core collapse under
self-gravity occurs in an inside-out manner; the `bounding
surface' (or the stagnation surface) of the magnetized collapsing
sphere travels outward at the magnetosonic speed (not a constant
in general) in a self-similar fashion. Shu (1977) utilized the
isothermal version of this EWCS solution to model the core
collapse of a molecular cloud in protostar formation. Zhou et al.
(1993) attempted to test Shu's isothermal dynamical model by
fitting the observed molecular spectral line profiles. However, it
should be noted that by incorporating a non-isothermal temperature
profile inferred from data, this isothermal model test by fitting
molecular spectral line profiles is not fully self-consistent. In
contrast, with a general polytropic model and the ideal gas law,
the temperature variation is a natural consequence. It would be
highly desirable to carry out a parallel analysis similar to that
of Zhou et al. (1993) but for a general polytropic gas cloud model
so that profiles of temperature, density and flow speed are all
self-consistently prescribed. By setting $h=0$ for the absence of
magnetic field, our generalized polytropic EWCSs are the same as
those of Cheng (1978).
%{\bf Need to check details.}
The outer magnetized SPS involves a completely random magnetic
field and thus a global quasi-spherical symmetry. As a result,
small-scale random flows with a zero mean are expected. In this
sense,
%we assumed that major inflow/outflow, i.e., the dominance of
%radial flow against transverse flows, the SPS here may be
%inconsistent because of lacking the radial flow. Thus,
the outer magnetized SPS portion here provides a large-scale mean
profile.
%{\bf However, because in our analysis on the magnetic field,
%major inflow/outflow should exist in order that the description
%on magnetic field is valid, the SPS here can only provide a role
%as an intuitive guide. Need to add some essential comments. }
%{\it to YQL: done with rewriting this sentence.}
The presence of a magnetic field would make a cloud to behave more
fluid like on large scales (the solar wind is an example) and can
give rise to radiative diagnostics when relativistic electrons are
involved.

%{\bf Degree of Freedom.}
The main difference between our general polytropic MHD EWCS here and
former EWCSs lies in the degree of freedom. For every determined MHD
flow with a set of parameters $\{\gamma,\ n,\ h\}$, there is at most
one EWCS and these three parameters are now allowed to change
independently. In the isothermal EWCS case of Shu (1977), we have
$\gamma=1$, $n=1$ and $h=0$.
%{\bf Please discuss the polytropic EWCS
%case of Cheng (1978). }{\it to YQL: done.}
In the conventional polytropic EWCS case of Lou \& Wang (2006), we
have $\gamma+n=2$ ($q=0$) and $h=0$. In the conventional
polytropic MHD problem studied by Wang \& Lou (2007), $q=0$ or
$n+\gamma=2$ is required and there are two parameters $\gamma$ and
$h$ that are allowed to vary independently. In the present
generalized polytropic MHD model, the parameter $n$ is also
allowed to change and is the key extension of this paper for this
model problem. Therefore, all polytropic MHD solutions in this
analysis have one more degree of freedom for fitting observational
data.
\begin{figure}
\includegraphics[width=3.3in,bb=100 160 500 630]{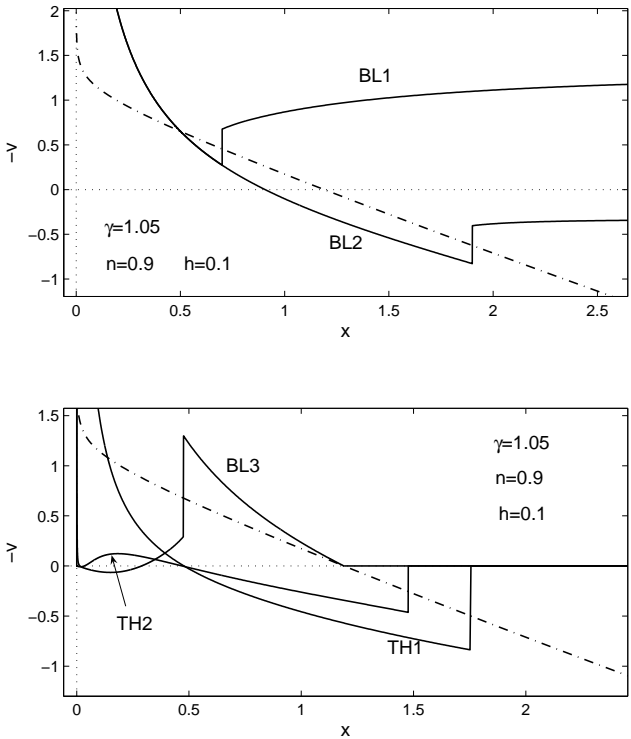}
\caption{
%{\bf You can place parameters in the figure.}
Semi-complete MHD shock solutions with free falls as $x\rightarrow
0$ with four parameters $\gamma=1.05,\ n=0.9,\ q=-0.143$, and
$h=0.1$.
%The two perpendicular light dashed straight lines are abscissa
%and ordinate axes. The dash-dotted curves are the magnetosonic
%critical curves.
In the upper panel, the MHD shock solution labelled with `BL1' is
constructed by integrating from a magnetosonic critical point
$(x,v,\alpha)=(0.5,-0.6520,3.053)$ inwards to small $x$ and outwards
to a MHD shock point
$(x_{s2},v_{s2},\alpha_{s2})=(0.7,-0.2731,2.191)$ downstream; using
MHD shock conditions $(\ref{shock6})-(\ref{shock8})$, we obtain the
corresponding upstream physical variables and thus integrate from
$(x_{s1},v_{s1},\alpha_{s1})=(0.70015,-0.6761,1.515)$ outwards to
large $x$. The `BL2' MHD shock solution is constructed by the same
inner portion with a different MHD shock point at
$(x_{s2},v_{s2},\alpha_{s2})=(1.9,0.8275,0.8481)$; again using MHD
shock conditions $(\ref{shock6})-(\ref{shock8})$, we determine the
upstream physical parameters at
$(x_{s1},v_{s1},\alpha_{s1})=(1.90061,0.4034,0.5728)$ and then
integrate outwards to large $x$. The `BL1' MHD solution has a large
$x$ asymptotic solution with $A=0.6536$ and $B=-1.780$
%(an inflow at large $x$)
in expressions (\ref{asym8}) and (\ref{asym9}); while these
parameters for `BL2' MHD shock solution are $A=2.168$ and $B=0.4534$
%(an outflow at large $x$)
in expressions (\ref{asym8}) and (\ref{asym9}). In the lower panel,
the `TH1' and `TH2' MHD solutions both have the same outer SPS
portion with $v=0$ and $\alpha=A_0x^{-2/n}$ as described by equation
(\ref{qstatic1}); this is the outer part of a magnetized SPS with
the radial profile of a random magnetic field given by relation
(\ref{transform5}). The MHD solution labelled with `TH1' has
upstream and downstream MHD shock points
$(x_{s1},v_{s1},\alpha_{s1})=(1.7555,0,0.5311)$ and
$(x_{s2},v_{s2},\alpha_{s2})=(1.7514,0.8358,1.130)$, and crosses the
magnetosonic critical curve at $(x,v,\alpha)=(0.1376,-1.081,12.20)$.
The MHD solution labelled with `TH2' has upstream and downstream MHD
shock points $(x_{s1},v_{s1},\alpha_{s1})=(1.4766,0,0.7800)$ and
$(x_{s2},v_{s2},\alpha_{s2})=(1.4760,0.4623,0.1963)$, and crosses
the magnetosonic critical curve at $(x,v,\alpha)=(7.825\times
10^{-6},-1.081,4.972\times 10^5)$. The `BL3' solution has shock
points $(x_{s1},v_{s1},\alpha_{s1})=(0.4758, -1.298, 3.289)$ and
$(x_{s2},v_{s2},\alpha_{s2})=(0.4743, -0.2907, 7.889)$, and crosses
the magnetosonic critical curve at
$(x,v,\alpha)=(4.885\times10^{-4},-1.945,4.474\times10^3)$. The
values of $m(0)$ for solutions `BL1', `BL2', `TH1', `TH2' and `BL3'
are 0.6987, 0.6987, 0.2665, 8.461$\times10^{-5}$ and
$2.589\times10^{-3}$. The values of shock parameter $\lambda$ for
these solutions are 1.000210, 1.000320, 1.00232, 1.000390 and
1.00303, respectively.
%{\bf Value of $\lambda$ for the case of BL3?}{\it to YQL: done}
%{\bf There are three nodes for `TH2' MHD shock solution?
%MHD shock solutions with even number of nodes can be also
%matched with a magnetized SPS!}
}\label{freefallshock}
\end{figure}

\subsection[]{Inner Free-Fall with Outer Inflow/Outflow}

Shu (1977) also constructed global semi-complete isothermal
solutions that approach a central free fall at small $x$, but with
$B=0$, $h=0$, $n=\gamma=1$ (or $q=0$) in asymptotic solution
(\ref{asym9}) and here $A>A_0$ for the outer asymptotic singular
isothermal sphere (SIS) solution (n.b., the singularity at $x=0$
is actually excluded in this construction). In fact, Shu used this
sequence of solutions to introduce the EWCS as the limiting case
and to suggest an inside-out collapse scenario for star formation.
As an example, we show a generalized polytropic MHD extension of
such solutions in Figure \ref{freefallnoshock}, labelled by `Shu2'
with $B=0$ and $A>A_0$; it is possible and straightforward to
construct a sequence of such solutions with $B=0$ by choosing
different $A$ values larger than $A_0$.
%{\bf You mean `Shu1'?}{\it to YQL: no, it is still Shu2, the Shu1
%has been discussed in the above subsection, this subsection is on
%the $B=0$ and $A>A0$ case.}
In addition, there is still the case of $B\neq 0$ in asymptotic
solution (\ref{asym9}), which is qualitatively similar yet has a
constant flow speed at large $x$ in the isothermal case for outer
envelopes (e.g., Lou \& Shen 2004; Shen \& Lou 2004; Fatuzzo et
al. 2004; Yu \& Lou 2005; Yu et al. 2006). There are increasing
observational evidence indicating that star forming clouds do have
systematic flows far away from the core (e.g., Fatuzzo et al.
2004) and our model can accommodate either inflows or outflows.

In general polytropic cases with $n<1$, the radial flow speed
remains finite and approaches zero at large $x$. With $n=1$, the
radial flow speed remains finite and approaches a constant value
$B$ at large $x$. While for $n>1$, the asymptotic flow term
associated with this $B$ coefficient in asymptotic solution
(\ref{asym9}) diverges at large $x$; to model a real system, one
then needs to set $B$ equal to zero. This parameter $B$ represents
a component of radial flow speed which satisfies the momentum
equation as if there is no force in action. Because Hunter (1977)
first constructed global semi-complete isothermal solutions with
this behaviour, we label `Hunter1', `Hunter2' and `Hunter3' in
Fig. \ref{freefallnoshock} as examples of this type of flow
solutions.
%{\bf Are you sure of the above Hunter statement and the
%no force in action interpretation?}{\it to YQL: yes.}
Here the `Hunter2' solution is one with a less dense envelope than
the SPS yet with an inflow speed; the `Hunter3' solution is one
with a more dense envelope yet with an outflow speed. Both involve
free-fall solutions in the central core collapse region.

Qualitatively, this type of general polytropic MHD solution has
either an inflow or an outflow in the outer envelope and a
free-fall inner core (Lou \& Shen 2004; Shen \& Lou 2004; Yu \&
Lou 2005; Yu et al. 2006; Lou \& Gao 2006). As the presence of
inflows is inferred observationally in star-forming regions and
molecular cloud cores (see, e.g., Fatuzzo, Adams \& Myers 2004 and
references therein), our model analysis here is more general in
the following four aspects, namely, (i) the inclusion of a
completely random magnetic field, (ii) the construction of an MHD
shock, (iii) the possibility of a constant `specific entropy'
everywhere at all time, and (iv) the possibility of `specific
entropy' conservation along streamlines yet with a variable
`specific entropy' in space and time. Physically, unless for a
`black hole' at the centre, the central free-fall of gas will
ultimately lead to a `sphere of transient activities' around the
core region; the change of equation of state and/or the ignition
of thermal nuclear reaction mark the onset of a star formation
process. In other words, we expect that the mathematical
singularity of a free-fall MHD solution as $x\rightarrow 0^+$ will
be taken care of by other relevant physical processes when a proto
star forms.
%is meant to be stopped by the change of equation of state
%when a star forms, so the singularity here is not physical.

%{\bf Degree of Freedom.}
Apart from the index parameters $n$ and $\gamma$ that are allowed
to vary independently for different polytropic MHD flows, there
are still two parameters $A$ and $B$ in asymptotic MHD solutions
(\ref{asym8}) and (\ref{asym9}). In a continuous manner, these two
parameters can change and be mapped to a central mass point $m(0)$
or central mass accretion rate. This mapping can be readily
determined by numerical integration. Fatuzzo et al. (2004) made a
semi-analytical estimate of this mapping.
%{\bf and Fatuzzo, Adams \& Myers (2004) made an effort to give
%a theoretical estimation. ?}{\it to YQL: rewriting done.}
In general, larger $A$ values or smaller $B$ values will lead to a
larger $m(0)$ as expected on intuitive ground. Observationally, this
corresponds to various central mass accretion rates in protostar
forming clouds.

\subsection[]{Envelope Expansion with Core\\
\quad\ Collapse in Magnetized Clouds}

Lou \& Shen (2004) constructed isothermal EECC solutions using a
solution matching technique in the $\alpha-v$ phase diagram (see
also Bian \& Lou 2005 and Wang \& Lou 2007). We constructed such
kind of general polytropic MHD solutions and present the first two
of this series of MHD solutions in the lower panel of Figure
\ref{freefallnoshock}. The first EECC solution labelled with `LS1'
is a MHD solution with an inner free fall (i.e., magnetized core
collapse) and an outer outflow (i.e., magnetized envelope
expansion). This kind of evolution behaviour may be qualitatively
applicable in asymptotic giant branch (AGB) stars, post-AGB stars,
and/or protoplanetary nebulae (PPNe); along this line, Lou \& Shen
(2004) proposed to utilize EECC solutions (including the second
`LS2' solution which is qualitatively different)
%{\bf (including the second solution which is qualitatively
%different)?  you mean `LS2' solution?}{\it to YQL: yes}
to grossly catch key features of MHD collapses and flows. Also in
the context of a collapsing magnetized cloud to form a protostellar
core, general polytropic MHD EECC solutions with $q\neq 0$
exemplified here appear more general to account for various
plausible situations (Shen \& Lou 2004; Lou \& Gao 2006). In all
these considerations, the inner singularity as $x\rightarrow 0^+$
can be removed once a departure from self-similarity occurs and/or a
different equation of state is adopted.

%{\bf Degree of Freedom.}
Except for the additional scaling parameter $n$, for every fixed MHD
flow profile there exists at most a series of discrete solutions,
each with a qualitatively different manner (e.g., different times to
cross the $v=0$ axis). It is common to this kind of flows (see also
Hunter 1977; Lou \& Shen 2004; Bian \& Lou 2005) that the $\alpha$
versus $v$ phase diagram obtained by integrating from the inner
portion (from $x$ values smaller than the meeting point $x_F$) tends
to spiral into a specific point. This point is on the phase curve
obtained by integrating from the outer portion (from $x$ values
larger than the meeting point $x_F$).
%{\bf This last sentence appears awkward. I can sort of
%guess what you mean. }{\it to YQL: rearranged, hope OK.}
\begin{figure}
\includegraphics[width=3.3in,bb=100 200 460 580]
{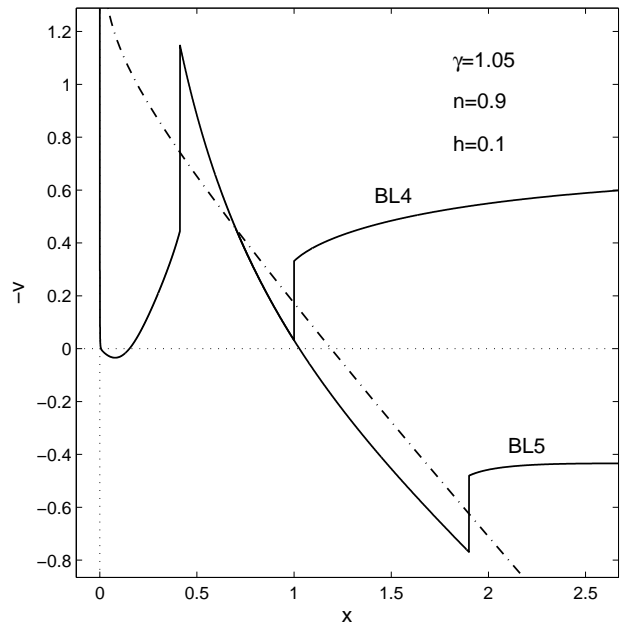}%-40,-30,+20,+10
\caption{Semi-complete self-similar polytropic solutions with twin
MHD shocks for parameters $\gamma=1.05$, $n=0.9$, $q=-0.143$,
$h=0.1$.
%The horizontal and vertical light dashed straight
%lines are abscissa and ordinate axes, respectively.
The dash-dotted curve is the magnetosonic critical curve. The inner
portion is the same for the two MHD solutions labelled with `BL4'
and `BL5', i.e., the MHD shock points for the inner shock are
$(x_{s2},v_{s2},\alpha_{s2})=(0.4129,-0.4458,6.949)$ for the
downstream side and
$(x_{s1},v_{s1},\alpha_{s1})=(0.4133,-1.147,3.743)$ for the upstream
side, and both MHD solutions also smoothly cross the magnetosonic
critical curve twice at the same two points at
$(x,v,\alpha)=(1.425\times10^{-4},-2.164,2.126\times10^4)$ and at
$(x,v,\alpha)=(0.7,-0.4543,2.146)$. The common inner portion of both
MHD solutions has two nodes (i.e., $v=0$) for $x<0.3$ and diverges
as $x\rightarrow 0$. Parameter $\lambda$ for the inner MHD shock is
1.00100 and parameter $m(0)$ for the central free-fall solution is
$9.338\times10^{-4}$. The MHD solution labelled with `BL4' has the
outer MHD shock at $(x_{s2},v_{s2},\alpha_{s2})=(1,-0.03170,1.512)$
for the downstream side and
$(x_{s1},v_{s1},\alpha_{s1})=(1.0000961,-0.3316,1.144)$ for the
upstream side; while the `BL5' MHD solution has the outer MHD shock
at $(x_{s2},v_{s2},\alpha_{s2})=(1.9,0.7693,0.8311)$ for the
downstream side and
$(x_{s1},v_{s1},\alpha_{s1})=(1.90019,0.4803,0.6358)$ for the
upstream side. The values of $\lambda$ for the two outer MHD shocks
of `BL4' and `BL5' are 1.0000961 and 1.000102. The `BL4' MHD
solution has a large $x$ asymptotic solutions (\ref{asym8}) and
(\ref{asym9}) with $A=1.076$ and $B=-0.9545$ (inflow), while for the
`BL5' MHD solution, the two corresponding parameters are $A=2.403$
and $B=0.6354$ (outflow).}\label{twinshock}
\end{figure}

\subsection[]{Expansion of MHD Shocks into an Outer Static Envelope Configuration}

Tsai \& Hsu (1995) constructed an isothermal self-similar shock
solution which connects an outer static configuration to a central
free-fall solution with a shock.
%{\bf Should also discuss champaign flow of Shu et al. (2002).}{\it to YQL:
%does not exist because Hunter type asymptotic solutions do not exist.}
Bian \& Lou (2005) further constructed various possible isothermal
shock solutions in a more comprehensive manner. We present here the
first three of such MHD shock solutions in the lower panel of Figure
\ref{freefallshock}.
%{\bf Another example is expected.}{\it to YQL: done.}
Based on extensive numerical exploration, Tsai \& Hsu (1995) also
suggested that in a star-forming cloud, when a strong burst of
thermal nuclear energy is released from the core, instead of a
smooth evolution in the form of EWCS as discussed by Shu (1977), a
shock can gradually emerge and travels outward in a self-similar
manner. Due to a sharp density profile of the singular isothermal
sphere (the counterpart of which is SPS in our model analysis here),
the shock travels long and evolves into a self-similar shock. They
provided numerical simulations to support this scenario of shock
initiation and evolution.
%Again, the SPS here is only an intuitive description, because
%major inflow/outflow is an essential assumption for our
%treatment on magnetic field. {\bf What is the key point here?}
%{\it to YQL: explained in detail in former comments on this point.}

%{\bf Degree of Freedom.}
The degree of freedom of this series of generalized polytropic MHD
shock solutions is like that of the MHD EECC solutions. Because
the outer SPS is prescribed, the MHD shock solutions need to be
properly matched in the $\alpha$ versus $v$ phase diagram. Similar
spiral-in features in the $\alpha$ versus $v$ phase diagram (e.g.,
Lou \& Shen 2004) may exist, leading to a series of discrete  MHD
shock solutions. Corresponding to each of such MHD shock
solutions, there is a specific central mass accretion rate.

%As the Larson-Penston type solutions at small $x$ no longer exist
%for polytropic flows with $q\neq 0$, the polytropic MHD
%counterpart of the so-called `champaign flows' studied by Shu et
%al. (2002) cannot be constructed here.
%
%Be careful with these statements. Larson-Penston type solutions at small $x$
%may exist for discrete values of $q\neq 0$.

\subsection[]{Generalized Polytropic MHD Shock\\
\quad\ \ Solutions with Inner Free Fall}

Bian \& Lou (2005) explored various isothermal hydrodynamic shock
solutions and noted astrophysical applications to AGB, PNe,
protostar formation, quasars and supernova explosions and so
forth. We here present two generalized polytropic MHD shock
solutions `BL1' and `BL2' as illustrative examples in Figure
\ref{freefallshock}. Within the framework of our model analysis,
it is possible to construct a variety of general polytropic MHD
shocks adapted to various astrophysical flow situations.

On the basis of a conventional polytropic hydrodynamic shock model
with $n+\gamma=2$ and thus $q=0$, Lou \& Gao (2006) examined
observationally inferred information for star-forming cloud cores.
They noted that star-forming regions are well studied in the inner
core and outer edge regions, which may be utilized to constrain or
test self-similar core collapse scenario with or without inflows
and/or outflows (Shen \& Lou 2004; Fatuzzo et al. 2004). While
globally smooth solutions such as EECC solutions have less degrees
of freedom when they encounter the sonic critical curve (which
occurs often), the possible existence of shocks is physically
plausible and mathematically convenient to join different inner
and outer asymptotic solutions into a global shock flow solution
(i.e., by crossing the sonic critical curve with a shock).
%{\bf What is done in this paper in context of star formation?}
%{\it to YQL: the main point is the extension of degree of
%freedom. Major physical interpretations are the same.}
Lou \& Gao (2006) also compared polytropic model with observations.
By taking into account of radiative transfer, we shall further fit
various observed molecular spectral line profiles for given
underlying general polytropic MHD shock flows with $q\neq 0$. Thus
shock solutions provide a simple and direct model scenario for
star-forming cloud cores that appear grossly quasi-spherical on
large scales.

%{\bf Degree of Freedom.}
Once parameters $\gamma$, $n$ and $h$ are prescribed, a global
semi-complete MHD shock solution has two degrees of freedom,
namely, one can choose the location where the solution smoothly
crosses the magnetosonic critical curve, and one can also choose
the MHD shock point; this is illustrated in the two panels of
Figure \ref{freefallshock} where two MHD shock solutions have
different outer envelopes. The location of smoothly crossing the
magnetosonic critical curve determines the parameter $m(0)$
(related to the central mass point and mass accretion rate) in the
core
%first {\bf region; quadrant?}{\it to
%YQL: `in the core' might be OK},
and the complete choice corresponds to a specific set of
parameters $A$ and $B$ for the outer large $x$ asymptotic MHD
solution (\ref{asym8}) and (\ref{asym9}). The plethora of this
type of MHD shock solutions can accommodate various astrophysical
situations including protostar-forming cloud cores.

\subsection[]{Twin Polytropic MHD Shock Solutions}

Bian \& Lou (2005) also constructed isothermal hydrodynamic twin
shock models, in which two shocks appear in a self-similar flow
and the solutions cross the sonic critical curve thrice, namely,
once smoothly and twice with shocks. For general polytropic MHD
flows, we present the counterpart solutions in Figure
\ref{twinshock}. This type of general polytropic MHD shock
solutions further expand the solution space, providing more
plausible models of quasi-spherical processes involving random
magnetic field, gravitational core collapse and far-away
inflows/outflows. We note that this general polytropic MHD twin
shock model differs from the so-called forward-reverse shock pair
(see, e.g., Chevalier 1982), because both shocks here are `forward
shocks' in the sense that the shock moves forward relative to the
local MHD flow. Conceptually, we emphasize the possibility of twin
or multiple general polytropic MHD shocks in a magnetized flow
system with $q\neq 0$ (see Yu et al. 2006 for the isothermal
case).

%{\bf Degree of Freedom.}
%The variety of this kind of polytropic MHD shock solutions
%is as many as the general polytropic MHD shock solutions.
When three parameters $\gamma$, $n$ and $h$ are chosen, one can
still adjust the place where the MHD flow crosses the magnetosonic
critical curve smoothly, and then choose the place where an outer
MHD shock appears; the former choice gives the central mass point
$m(0)$ or central mass accretion rate, while for the latter
choice, the shock location $x$ also corresponds to the outward
travel speed of this MHD shock. Figure \ref{twinshock} shows that
different choices of outer MHD shocks correspond to very different
outer MHD flow profiles in the envelope.

\subsection[]{Inner Quasi-Magnetostatic\\
\quad\ \ Solutions with Outer Inflows}

\begin{figure}
\includegraphics[width=3.3in,bb=120 220 460 580]{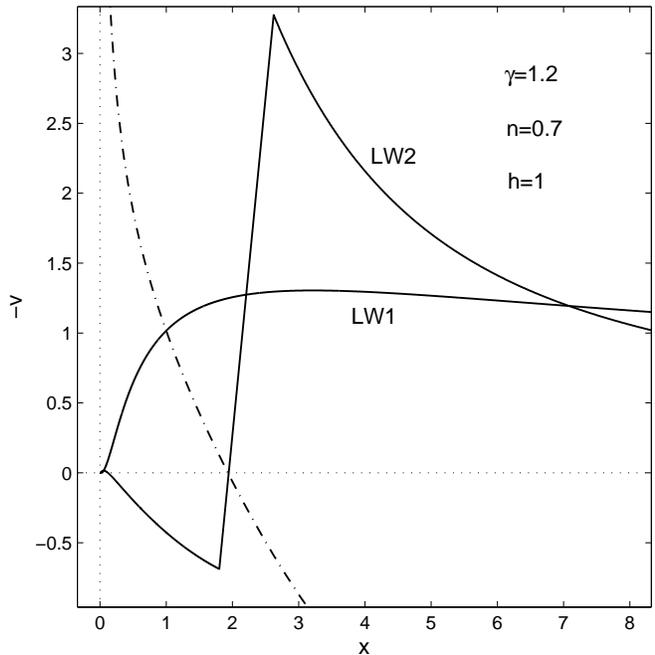}
\caption{Semi-complete solutions with quasi-magnetostatic asymptotic
solutions (\ref{qstatic2}) at small $x$ for $\gamma=1.2$, $n=0.7$,
$q=-2$ and $h=1$. We then have $Q=-3.308$ by definition
(\ref{qstatic5}) and $A_0=1.078$ in expression (\ref{qstatic2}).
%The two perpendicular light dotted straight
%lines are abscissa and ordinate axes.
The dash-dotted curve is the
magnetosonic critical curve. In this case, the parameter $K$
obtained from quadratic equation (\ref{qstatic6}) for the
quasi-magnetostatic solution (\ref{qstatic2}) is
$K=K_1+iK_2=1.357+0.8341i$. The MHD solution labelled with `LW1' is
constructed by starting from $(x,v,\alpha)= (1,-1.016,0.3583)$ on
the magnetosonic critical curve and integrating both inwards and
outwards numerically. For the inner quasi-magnetostatic asymptotic
solution at small $x$, parameter $L$ is $L=-0.02262-0.8951i$, and
thus $N=-0.2365-1.459i$; while for the outer large $x$ MHD
asymptotic solution (\ref{asym8}) and (\ref{asym9}), we have
corresponding parameters $A=0.7662$ and $B=-3.314$ for an inflow.
The polytropic MHD shock solution labelled with `LW2' decribes a
rebound MHD shock in a self-similar evolution (see Lou \& Wang 2007
for more details); this solution is constructed as follows. We start
by integrating from $(x,v,\alpha)= (3,\ 0.8699,\ 0.1134)$ on the
magnetosonic critical curve inwards and then stop at
$(x_{s2},v_{s2},\alpha_{s2})=(1.8,\ 0.6877,\ 0.5869)$ which is
regarded as the downstream side of an MHD shock and this segment of
solution is itself ignored. Using MHD shock conditions
(\ref{shock6})$-$(\ref{shock8}), we then derive physical variables
$(x_{s1},v_{s1},\alpha_{s1})=(2.623,-3.275,0.09576)$ upstream of the
shock. Finally, we integrate inwards from the downstream side with
$(x_{s2},v_{s2},\alpha_{s2})=(1.8,\ 0.6877,\ 0.5869)$ and outwards
from the upstream side with $(x_{s1},v_{s1},\alpha_{s1})
=(2.623,-3.275,0.09576)$. The parameter $\lambda$ for this
polytropic MHD rebound shock is $\lambda=1.457$ and the outer
portion has a large $x$ asymptotic MHD solution with parameters
$A=3.890$ and $B=-1.525$ for an inflow.
%{\bf It would be helpful for reference to also
%provide specific values of $N$, $Q$ and $A_0$
%for these two examples.} {\it to YQL: done}
The inner quasi-magnetostatic solution has parameter
$L=0.9797+0.4042i$ and thus $N=1.692+0.4427i$. Both MHD solutions
oscillate in the small $x$ regime in a self-similar manner (Lou \&
Wang 2006; Wang \& Lou 2007). }\label{quasistatic}
\end{figure}

Using a conventional polytropic hydrodynamic formulation with $q=0$,
Lou \& Wang (2006) first presented the quasi-static asymptotic
solution in the regime of small $x$ and then constructed a rebound
shock model to catch certain gross features of a class of
supernovae. We attempted to relate the mass of a progenitor star and
the type of a remnant compact object (such as a white dwarf, a
neutron star or a black hole) left behind after a gravitational core
collapse and a subsequent emergence of rebound shock.
%{\bf A discussion of Lou \& Wang 2007 here?}
The above model can be extended to include a completely random
transverse magnetic field with quasi-spherical symmetry (Lou \&
Wang 2007). In this MHD rebound shock model, we explore the
origin(s) of intense magnetic fields on compact objects from the
perspective of fossil field associated with massive progenitor
stars. As an important part of further model development, we have
also constructed generalized polytropic MHD solutions that
smoothly cross the magnetosonic critical curve and approach this
quasi-magnetostatic asymptotic solution (\ref{qstatic2}) as shown
in Figure \ref{quasistatic} and labelled with `LW1'. In the regime
of small $x$ and with complex coefficients $K$ and $L$, the MHD
flow velocity actually oscillates with a decreasing magnitude as
$x\rightarrow 0^+$ (not easily seen here; see Lou \& Wang 2006 for
a similar example in detail). Qualitatively speaking, this
generaized polytropic MHD solution represents a far-away or
initial inflow leading to an eventual quasi-magnetostatic inner
core and carries a desirable feature of forming a central
magnetostatic configuration instead of a divergent free fall
towards a central mass point. Therefore, this type of generalized
polytropic MHD solutions should also provide a physically
plausible model framework to study protostar formation processes
(Shen \& Lou 2004; Lou \& Gao 2006); we shall discuss this
important application in separate papers.

%{\bf Degree of Freedom.}
This type of general polytropic MHD solution has one degree of
freedom in construction, namely, the choice of the point where a
solution crosses the magnetosonic critical curve.

\subsection[]{Inner Quasi-Magnetostatic Solutions\\
\quad\ \ with Rebound MHD Shocks}

Lou \& Wang (2006) and subsequently Lou \& Wang (2007) proposed a
conventional polytropic gas model (i.e., $q=0$ and thus
$n+\gamma=2$) for rebound shock process in a class of supernova
explosions. In particular, Lou \& Wang (2007) have incorporated a
random magnetic field (e.g., Zel'dovich \& Novikov 1971) and
attempted to address the fundamental issue on the physical origin
of strong magnetic field of a compact object left behind a rebound
MHD shock. Based on our estimates, we proposed that for a
progenitor star of initial mass in the range of $\sim
6-8M_{\odot}$, an MHD rebound shock initiated during the core
collapse may eventually leave behind a magnetic white dwarf with
magnetic field strengths in the range of $\sim 10^6-10^8$G
depending on surface magnetic field strengths of the progenitor
star. Similarly, for a progenitor star of initial mass in the
range of $\gsim 8M_{\odot}$, a stronger MHD rebound shock
initiated during the gravitational core collapse may ultimately
leave behind a neutron star with magnetic field strengths in the
range of $\sim 10^{11}-10^{12}$G depending on surface magnetic
field strengths of progenitor stars. In reference to the key
result of Lou \& Wang (2007), the generalized polytropic MHD model
here allows $n+\gamma\neq 2$ and thus $q\neq 0$. The major
physical consequence is that `specific entropy' is conserved along
streamlines but `specific entropy' is not uniformly distributed in
space and time during the gravity-induced core collapse and
rebounce within a progenitor star. As an example of illustration,
we display our generalized polytropic rebound MHD shock solution
labelled with `LW2' in Figure \ref{quasistatic}. This solution
carries a feature that a strong MHD shock emerges surrounding an
inner magnetized core, ploughs through an infalling magnetized
outer envelope (either with a stellar wind or just with an inflow)
and leaves behind a quasi-magnetostatic compact configuration of
high density. The `rebound' MHD shock here is a physical
simplification of a neutrino-driven shock, as opposed to the
earlier concept of a `prompt' shock.
%which is a concept that is obsolete from modern point of view.
One important flexibility here is that the constraint $n+\gamma=2$
is now relaxed such that relevant coefficients allow for various
plausible combinations. In the analysis of Lou \& Wang (2007), we
have already pointed out that the asymptotic scaling laws of the
mass density profile and magnetic field profile in terms of radius
$r$ is independent of the equation of state, which is apparent from
the analysis of this paper. This is the crucial difference between
this model under the present formulation and those of former
analyses. For example, in our former model (Lou \& Wang 2007), it
appears that
\begin{equation}\label{add1}
<B^2_t>^{1/2}\sim r^{1-2/n},
\end{equation}
and we need $n$ to approach $2/3$ for the strongest magnetic field
amplification or the fastest field variation for the stellar
surface to the central core; this appears to constrain the value
of polytropic index $\gamma$ in our former analysis with $q=0$.
However, in the current model framework with $q\neq 0$, variations
of $n$ and $\gamma$ are no longer constrained in this regard. We
can still let $n$ approaches $2/3$ but the $\gamma$ value is not
necessarily close to $4/3$.
%not a limitation to the equation of state.}
%{\bf Please clarify this point more explicitly.}

%{\bf Degree of Freedom.}
In constructing such general polytropic rebound MHD shock solution,
we can choose the point where the solution would cross the
magnetosonic critical curve smoothly (see Fig. \ref{quasistatic}
caption for the solution construction procedure; the solution
portion between the magnetosonic critical point and the downstream
MHD shock point is actually ignored later) and the MHD shock point.
Therefore except for the choice of parameters $\gamma$, $n$ and $h$,
we still have two degrees of freedom for constructing such type of
semi-complete solutions with $q\neq 0$.

To describe a collapsing stellar core prior to the emergence of a
rebound shock for a supernova explosion, Goldreich \& Weber (1980)
derived an exact homologous solution using a polytropic model with
$\gamma=4/3$ and constant specific entropy everywhere within the
stellar core, compared with numerical simulations (Van Riper \&
Arnett 1978; Bethe et al. 1979), and identified parameter range
allowing for a homologous core collapse. In the light of current
model analysis and complementary to the analysis of Goldreich \&
Weber (1980), Lou \& Cao (2008) realized that it is not necessary
to impose the condition of a constant specific entropy throughout
the stellar core (also unlikely in reality) and one can still
derive homologous solutions for a collapsing stellar core.
Besides, we further obtain a broad class of self-similar solutions
for $\gamma=4/3$ with or without shocks.

\subsection[]{Inner Magnetoaccretion with Outer Inflows}

\begin{figure}
\includegraphics[width=3.3in,bb=120 200 460 580]{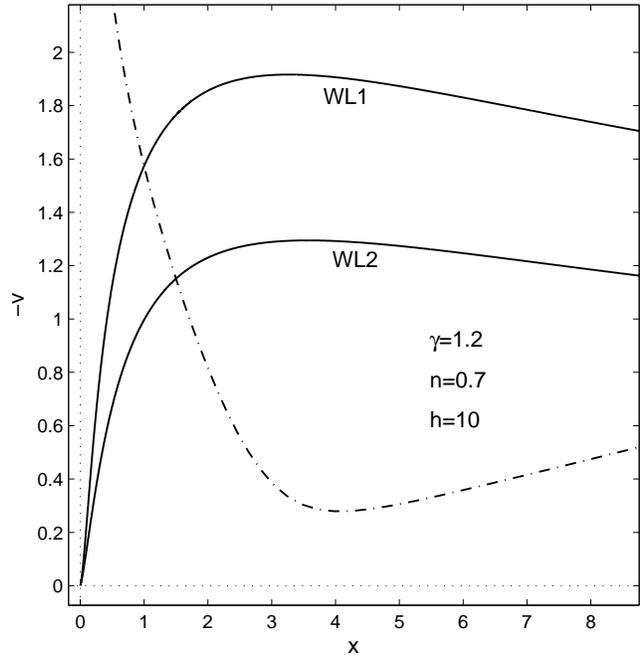}
\caption{Semi-complete polytropic MHD solutions with inner
magnetoaccretion asymptotic solution (\ref{asym5}) and (\ref{asym6})
in the strong-field regime with $\gamma=1.2$, $n=0.7$, $q=-2$ and
$h=10$.
%The horizontal and vertical light dotted straight
%lines are abscissa and ordinate axes, respectively.
The dash-dotted curve is the magnetosonic critical curve. For these
parameters, the value of the proportional coefficient of $v$ versus
$x$ is $-0.1873$ in solution (\ref{asym5}) for the upper plus sign
and this value is adopted for both WL1 and WL2 solutions.
%{\bf this value is for both WL1 and
%WL2 solutions?}{\it to YQL: yes.}.
The MHD solution labelled with `WL1' is constructed by integrating
from $(x,v,\alpha)=(1,-1.574,0.2671)$ on the magnetosonic critical
curve both inwards and outwards. Parameter $D_0$ in solution
(\ref{asym6}) is about $1.4$ and the outer portion has large $x$
asymptotic solution (\ref{asym8}) and (\ref{asym9}) with parameters
$A=0.7301$ and $B=-5.155$ (inflow). The `WL2' solution is
constructed similarly, but the starting point is
$(x,v,\alpha)=(1.5,-1.151,0.1452)$ on the magnetosonic critical
curve; parameter $D_0$ in solution (\ref{asym6}) is about $1.5$, and
the large $x$ MHD asymptotic solution (\ref{asym8}) and
(\ref{asym9}) has parameters $A=0.8738$ and $B=-3.494$ (inflow).
These solutions do not encounter the magnetosonic curve again. It is
also possible to construct MHD shock solutions in this scheme.
%{\bf Which sign of solution is selected? Do these solutions
%encounter the magnetosonic critical curve again?}
%{\bf There are two possible strong-field solutions in general for
%a given set of parameters. How about the other one? WWG indicates
%an unstable outward numerical integration.}
%{\bf Is it possible to construct corresponding MHD shock solution?}
%{\it to YQL: the one with a larger C value. It does not encounter the
%magnetosonic critical curve again. It is possible to construct shocked
%solutions, but currently we are not applying this solution to a particular
%astrophysical systems and a shock may seem to be a little weird.}
}\label{magnetoaccretion}
\end{figure}

Wang \& Lou (2007) report a novel magnetoaccretion solution for a
conventional polytropic gas ($q=0$) with a completely random
magnetic field. The isothermal MHD counterpart of this
magnetoaccretion solution was described in Yu et al. (2006). We
here construct the counterpart solutions (\ref{asym5}) and
(\ref{asym6}) with the upper signs in a generalized polytropic
($q\neq 0$) MHD formulation and display two examples of these
results in Figure \ref{magnetoaccretion}. These solutions involve
fairly strong magnetic field (i.e., $h>4$). Especially at small
$x$, we would expect a `sphere' of magnetic activities and
transients within which the self-similarity and quasi-spherical
symmetry are destroyed.
%It is possible that this solution can only be treated as
%a qualitative model, because the magnetic field is strong
%for this asymptotic solution.
However, the physical scenario corresponding to this solution, i.e.,
the possibility that the gravitational energy can be effectively
converted into magnetic energy via magnetoacretion processes, is
tantalizing. We shall further analyze this solution and its
astrophysical applications in separate papers.

%{\bf Degree of Freedom.}
In constructing this type of general polytropic MHD solution, we can
choose the point where the solution crosses the magnetosonic
critical curve smoothly and therefore we have only one degree of
freedom.
Numerical integrations with asymptotic solutions (\ref{asym5}) and
(\ref{asym6}) for the lower signs appear unstable in our experiment.
This problem remains to be investigated further (Lou \& Bai 2007 in
preparation).

\subsection[]{MHD Solutions of Inner Thermal\\
\qquad Fall and Outer Thermal Expansion}

\begin{figure}
\includegraphics[width=3.3in,bb=100 120 460 580]{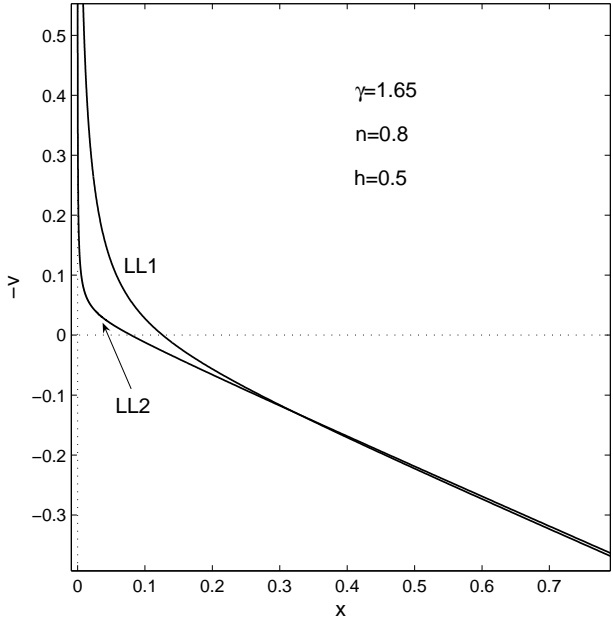}
\caption{Semi-complete MHD solutions with inner thermal fall
asymptotic solutions (\ref{thermalfall3}) and (\ref{thermalfall4})
at small $x$ and outer MHD thermal expansion asymptotic solutions at
large $x$ (see section 3.1.5) for
%the following parameters
$\gamma=1.65$, $n=0.8$, $q=2.25$ and $h=0.5$. The `LL1' and `LL2'
MHD solutions have parameter $m(0)=0.4$ and $m(0)=0.2$,
respectively, and are constructed by integrating from small $x$
outwards with above parameters. The outer portions of these two MHD
solutions converge to the same asymptotic solution at large $x$ to
the first order, with the proportional constant $c$ of $v(x)\sim cx$
being $0.5063$ for the thermal expansion solution and the
corresponding constant coefficient $E$ for this solution is $2.277$,
and the value of $P$ as the power law index for $\alpha(x)$ versus
$x$ is $-1.638$. By numerical experiment, these two different
solutions at small $x$ gradually merge to the same asymptotic
solution at large $x$ and do not encounter the magnetosonic curve.
%{\it to YQL: yes.}
%{\bf Please describe this MHD solution in details. For example,
%the procedure of construction and the value of parameter $P$ etc.
%Where is the magnetosonic critical curve?}{\it to YQL: For the
%critical curve problem please refer to the email.} {\bf Is it
%possible to construct construct MHD shock solutions?}{\it to
%YQL: no. Because it appears that there are no real root for the
%jump condition. Besides it is also weird if we do not apply these
%to a real astrophysical system.}
}\label{thermalfall}
\end{figure}

The hydrodynamic thermal fall asymptotic solution at small $x$
reported by Lou \& Li (2007 in preparation) is in a parameter
regime quite different from earlier analyses (i.e., $\gamma>4/3$
here). In our polytropic MHD generalization, this asymptotic MHD
solution (\ref{thermalfall3}) and (\ref{thermalfall4}) seems to
directly connect with the MHD thermal expansion solution
(\ref{thermalex1})$-$(\ref{thermalex5}) at large $x$ as presented
in Figure \ref{thermalfall}. While not being a proof, our
extensive numerical exploration seems to indicate that this type
of MHD solutions does not encounter the magnetosonic critical
curve.

%{\bf Need to clarify the situation of the magnetosonic critical
%curve in Fig. \ref{thermalfall}.} {\bf Are you sure that this type
%of solutions can never go across the magnetosonic critical curve?
%Reason? }

%{\bf Degree of Freedom.}
There is one degree of freedom for constructing this type of MHD
solutions, namely, the choice of inner mass point $m(0)$, once
parameters $\gamma$, $n$ and $h$ have been chosen.

\section[]{Discussion}

%{\bf Please discuss the special case of $\gamma=4/3$.}
%{\bf Please discuss the non-existence of exact global MHD solution.}
%{\bf Please discuss the non-existence of Larson-Penston type of
%solutions for $q\neq 0$ at small $x$ in a clear and more substantial
%manner. Implications of this dependence on $q$ parameter? }
In this paper, we have shown that more general polytropic
self-similar MHD shock flow solutions can be obtained under
various conditions and they can be applied or adapted to various
astrophysics systems by examples. In particular, we focus on the
case of $q\neq 0$, that is, the specific entropy is conserved
along streamlines but is not necessarily constant in space $r$ and
time $t$. More specifically, the specific entropy is related to
the enclosed mass that varies in both space and time in general.
While this more general polytropic MHD formalism extends the
parameter regimes of several known classes of self-similar
solutions, this same condition of $q\neq 0$ (or $n+\gamma\neq 2$)
also excludes certain solutions, notably
%the Larson-Penston type of solutions and
the exact global MHD solution (Wang \& Lou 2007). In astrophysical
applications, it would be informative and important to know these
generalizations and constraints for relevant theoretical model
development.

In this paper, we mainly studied quasi-spherical polytropic MHD
gas flows as models for various stellar level dynamical processes.
However, with various sensible extensions and adaptations (such as
the two-fluid problem with dark matter-baryon matter coupling by
gravity at the galaxy cluster level; Lou 2005; Lou, Jiang \& Jin
2008; Jiang \& Lou 2008 in preparation), similarity solutions can
be applied to different plasma flow systems with their own
characteristic spatial and temporal scales. Another example is the
attempt of Suto \& Silk (1988) to relate self-similar solutions to
galaxy formation. In principle, if an MHD system experiences a
long dynamic evolution, it may behave self-similarly and thus be
described by our model.
%Several of our future papers will focus on other
%astrophysical/cosmological problems on other degrees of magnitude.
We emphasize that in Suto \& Silk (1988), their cases of $n=1$
with different values of $\gamma\neq 1$ do not correspond to
specific entropy conservation along streamlines but correspond to
specific entropy variations in time $t$ only. In our cases of
$n=1$ with different values of $\gamma\neq 1$, the specific
entropy is a function of both $r$ and $t$ and, in particular, is
conserved along streamlines.

%{\bf
In the preceding section \ref{appl}, we have systematically
presented various astrophysical applications and implications of
our theoretical model development in reference to earlier
observations and theoretical models in relevant astrophysical
contexts. These include the following cases.

\noindent
(1) The isothermal EWCS (Shu 1977; Shu et al. 1987) for the
inside-out collapse scenario of star formation can now be
generalized by our model here in several aspects, namely, the
inclusion of a completely random magnetic field in the collapsing
cloud and the more general polytropic ISM without the restriction
of an isothermal gas. These two aspects provide the modelling
basis for massive star formation, radio synchrotron and/or x-ray
emission diagnostics and radiative synthesis of molecular line
profiles in star-forming clouds. In particular, a more realistic
non-isothermal temperature profile can be determined within a
self-consistent polytropic model framework. This is extremely
important for modelling protostellar cores embedded in molecular
clouds.

\noindent
(2) The isothermal EECC solutions (Lou \& Shen 2004; Shen \& Lou
2004) for the envelope expansion and core collapse scenario of
star formation in molecular clouds can now be generalized by our
model here in several aspects, namely, the inclusion of a
completely random magnetic field in the collapsing cloud and the
more general polytropic ISM without the restriction of an
isothermal gas. In addition to the implications already noted
immediately above in item (1), we can now take into account of
inflows (Fatuzzo et al. 2004), outflows of ISM in clouds as well
as MHD shocks in a more realistic manner. There are growing
observational evidence for inflows, outflows and shocks around
star-forming cores in ISM clouds. Our MHD shock flow solutions are
necessary for certain observed line profiles (Lou \& Gao 2008 in
preparation).

\noindent
(3) The isothermal magnetized EECC solutions (Yu \& Lou 2005; Yu
et al. 2006) is now replaced by a more general polytropic plasma
flow which is permeated with a completely random magnetic field.
By removing the isothermal constraint, the radial scaling of mass
density profiles in a cloud may vary from $\sim r^{-1}$ to
$r^{-3}$; observationally inferred radial density profiles of
molecular clouds do fall within this range. For the hot magnetized
intracluster medium (ICM) in the context of clusters of galaxies
on Mpc scales (e.g., Lou et al. 2008; Jiang \& Lou 2008 in
preparation), we predict the same mass density profile range.

\noindent
(4) The isothermal shock models (Tsai \& Hsu 1996; Shu et al.
2002; Shen \& Lou 2004) are substantially extended by our MHD
model here in several aspects, namely, the inclusion of a
completely random magnetic field, the specific entropy
conservation along streamlines, and the inclusion of inflows and
outflows. Our models are more versatile in modelling the so-called
``champagne flows" in star-forming H$_{\rm II}$ regions which
involve ionization fronts and shocks.

\noindent
(5) Various hydrodynamic isothermal shocks have been extensively
explored by Bian \& Lou (2005) for possible asymptotic flows at
large and small radii. For modelling ``champagne flows" in
star-forming H$_{\rm II}$ regions, Shu et al. (2002) considered
only central isothermal Larson-Penston type solutions. We can now
further construct ``champagne flows" with central free falls, with
a completely random magnetic field and with MHD shocks. It is also
possible to construct MHD shock flows associated with different
central solutions. In general, relativistic cosmic-ray particles
(electrons in particular) can be abundantly produced in either MHD
wind shocks or accretion shocks.

\noindent
(6) The quasi-static model solutions first obtained in a
conventional polytropic gas (Lou \& Wang 2006, 2007; Wang \& Lou
2007) are now generalized to more general polytropic magnetized
gas flows in this paper. Based on the new quasi-static asymptotic
solutions, Lou \& Wang (2006, 2007) proposed a new class of
rebound shocks for supernovae. Among other things, we would like
to understand the ejection of stellar materials of the progenitor
star, the mass of a compact object left behind, as well as the
origin of intense magnetic fields of compact objects. For a
rebound shock process in a progenitor star with the specific
entropy conserved along streamlines, the entropy distribution in
association with the enclosed mass becomes an important aspect for
various possible solutions (see also Lou \& Cao 2008 for the case
of $\gamma=4/3$). This can give rise to a variety of rebound MHD
shocks in supernova explosions. Of course, we still need to learn
about the specific energetic corresponding to any chosen entropy
distribution at a certain stage.

\noindent
(7) The asymptotic magneto-accretion solutions with magnetic
parameter $h>4$ were first studied by Yu \& Lou (2006) for the
case of an isothermal gas and by Wang \& Lou (2007) for a
magnetized conventional polytropic gas. This asymptotic MHD
solution has now been extended to a general polytropic MHD gas
flow and may be relevant to extremely intense magnetic fields
inferred for magnetars (e.g., $10^{13}-10^{15}$G). A strong
surface magnetic field of the progenitor star is implied. Magnetic
Ap stars have surface magnetic fields of several to ten thousand
gauss and give a typical $h>4$ by estimates.
%}

While our model results are mainly restricted\footnote{For $n>2/3$
and $\gamma>4/3$, we should proceed with care about magnetosonic
critical curves with $n+\gamma>2$.} to $n>2/3$
%{\bf should be $n>2/3$?}  wwg yes
and $\gamma<4/3$
%{\bf Why do we require $\gamma<4/3$? Need to clarify
%the situation of critical curve for $n+\gamma>2$.}
for most situations (except that the thermal fall and thermal
expansion solutions have $3/2<\gamma<5/3$), the special case of
$\gamma=4/3$ contains a variety of substantially new solutions
(Lou \& Cao 2008) that differ from those homologous core collapse
solutions explored earlier by Goldreich \& Weber (1980) in the
context of supernovae. Lou \& Cao (2008) describe this new
theoretical model development in the absence of a completely
random magnetic field and obtain various novel results as compared
to our analyses here. In particular, the constant coefficient
$\mathcal C$ in equation (\ref{add2}) cannot be absorbed by a
rescaling transformation, and features of hydrodynamic as well as
MHD systems will depend upon the choice of this constant
coefficient $\mathcal C$ in a nontrivial manner.

%Lou \& Li (2007 in preparation) reported that the LP solution (with
%$v=2x/3$ and $\alpha=$ constant) does not exist for $q\neq 0$. Here,
%we follow their analysis and find that the direct counterpart of the
%Hunter type asymptotic solution (with $v\sim 2x/3$ for small $x$)
%also does not exist for $q\neq 0$.
%The mathematical reason for this non-existence is that when
%$q=0$, the pressure force $\beta'$ in the radial momentum equation
%vanishes, while in the $q\neq0$ case it does not. It is interesting
%because for small $x$, $x^\xi$ is smaller than a constant (provided
%that $\xi>0$), but their derivatives are reversed, that is, the
%derivative of a constant vanishes while the derivative of $x^\xi$
%versus $x$ does not. Hence, a variety of semi-complete solutions
%having this asymptotic behaviour, such as those crossing the critical
%line twice (like EECC solutions, as constructed in Hunter 1977) and
%with shocks or twin shocks (such as those in Bian \& Lou 2005),
%disappear in our model. In particular, the well known `champaign
%flow' of Shu et al. (2002) also does not exist. This non-existence
%is interesting in the sense that $q=0$ case is not likely to occur
%exactly and thus how to explain this feature remains an open question.

Parallel to the analysis of Zhou et al. (1993),
%{\bf Please ask Gao Yang to give you the paper by Zhou S.D. et
%al. (1992?) on modelling spectral lines.}{\it to YQL: done.}
in which the isothermal model of Shu (1977) was claimed to be
supported by observational inferences, we are investigating
radiative diagnostics of the velocity, density and temperature
profiles in collapsing cores of clouds from molecular spectral
line profiles (Lou \& Gao 2008 in preparation). Meanwhile,
magnetic fields in proto-stellar clouds not only affect their
dynamic evolution, but also provide diagnostic signals, such as
synchrotron radiation and shock acceleration of relativistic
electrons etc. By including radiative transfer processes, our
model is capable of predicting such diagnostic signals.
%, which will be worked out and reported in a separate paper.

We also note earlier theoretical analyses on the instability of
known self-similar solutions (e.g., Ori \& Piran 1988; Hanawa \&
Nakayama 1997; Hanawa \& Matsumoto 1999, 2000; Semelin, Sanchez \&
de Vega 2001; Lou \& Bai 2008 in preparation). Instability
analysis helps to test whether certain self-similar flows will
actually occur in nature; or if they do occur, whether they will
last long enough. Our model provides a variety of new polytropic
MHD self-similar solutions with or without shocks, all ready to be
examined for such an instability analysis. At this stage, these
stability questions remain completely open.

Finally, numerical simulations are powerful and necessary in
determining how likely our various self-similar MHD solutions will
eventually emerge in a sensible way for plausible initial and
boundary conditions based upon astrophysical input. Moreover,
nonlinear instability analysis also needs to employ numerical
simulations. We hope that the results presented here will trigger
extensive numerical explorations for transient behaviours leading
to these self-similar MHD solutions with specific entropy
conserved along streamlines.

%{\bf The Discussion should be enriched if possible.}

%\section[]{Summary and Discussion}\label{disc}

\begin{acknowledgements}
This research has been supported in part by the ASCI Center for
Astrophysical Thermonuclear Flashes at the University of Chicago,
%under the Department of Energy contract B341495, by the Special
%Funds for Major State Basic Science Research Projects of China,
by the Tsinghua Center for Astrophysics, by the Collaborative
Research Fund from the National Science Foundation of China (NSFC)
for Young Outstanding Overseas Chinese Scholars (NSFC 10028306) at
the National Astronomical Observatories, Chinese Academy of
Sciences, by the NSFC grants 10373009 and 10533020 at the Tsinghua
University, and by the SRFDP 20050003088
%Specialized Research Fund for the
%Doctoral Program of Higher Education
and the Yangtze Endowment from the Ministry of Education at Tsinghua
University.
%The hospitalities of the Mullard Space Science
%Laboratory at University College London, U.K., of School
%of Physics and Astronomy, University of St Andrews, Scotland,
%U.K., and of Centre de Physique des Particules de Marseille
%(CPPM/IN2P3/CNRS) et Universit\'e de la M\'editerran\'ee
%Aix-Marseille II, France are also gratefully acknowledged.
%Affiliated institutions of Y-QL share this contribution.
\end{acknowledgements}

\appendix

\section[]{Comments on the Positive Definiteness of
\\ \quad \lowercase{$nx-v$}}\label{posdef}

%\section[]{Magnetic Induction and the Lorentz Force}\label{QSphere}

For a self-similar gas flow where the self-gravity may or may not
play a significant role
%{\bf Clarification needed.}
and for the self-similar transformation in the form of
\begin{equation}\label{posdef1}
r=k^{1/2}t^nx\ ,\quad u=k^{1/2}t^{n-1}v\ , \quad \rho=\alpha/(4\pi
Dt^s)\ ,
\end{equation}
%{\bf transformation not complete here?! Please define notations
%explicitly. }{\it to YQL: Yes, only these three are needed.}
where $D$ and $k$ are two constant positive parameters and $n$ and
$s$ are two exponents with $3n-s>0$, we then derive from mass
conservation equation (\ref{basic1})
\begin{equation}\label{posdef2}
M(r_2)-M(r_1)=\frac{k^{3/2}t^{3n-s}}{(3n-s)D}\alpha
x^2(nx-v)\bigg|^{x_2}_{x_1}\ ,
\end{equation}
where $x_i=r_i/(k^{1/2}t^n)$ with subscript $i=1,\ 2$ at a given
time $t$. The condition for equation (\ref{posdef2}) to be valid is
that the gas flow within the radial range between $r_2$ and $r_1$
evolves in a self-similar manner. Equivalently, we can express the
total enclosed mass $M(r)$ as
\begin{equation}\label{posdef3}
M(r)=\frac{k^{3/2}t^{3n-s}}{(3n-s)D}\alpha
x^2(nx-v)\bigg|^{r/(k^{1/2}t^n)}_{r_0/(k^{1/2}t^n)}+M(r_0)\ ,
\end{equation}
as long as the gas flow evolves in a self-similar manner within the
radial range between $r$ and $r_0$.

%{\bf need a bit more explanations!}{\it to YQL: done. Hope OK}
In general, $nx-v>0$ is not necessarily required because $M(r_0)$
may not behave self-similarly (in time or in space); thus the only
requirement for equation (\ref{posdef3}) is that $\alpha x^2(nx-v)$
is a monotonically increasing function of $x$, which is guaranteed
by the mass conservation equation, provided that $\alpha>0$ and
$x>0$. This is why in self-similar flow systems where the
self-gravity is not important (e.g., Chevalier 1982), solutions with
$nx-v<0$ are allowed (e.g., for the reverse shock portion of the
solutions in Chevalier 1982).

In contrast to the above case, we now consider the situation where
the self-gravity is indeed important. If we are to establish a
self-similar evolution, the total enclosed mass $M(r,t)$ would enter
the radial momentum equation and thus be a self-similar variable
under consideration. From a dimensional analysis, if we take only
$k$ and $D$ as dimensional parameters, the sensible transformation
for total enclosed mass $M(r,t)$ in the power law form should be
defined as
\begin{equation}\label{posdef4}
M=\frac{k^{3/2}t^{3n-s}}{(3n-s)D}m(x)\ ,
\end{equation}
where the reduced enclosed $m(x)$ depends on $x$ only. Comparing
equations (\ref{posdef3}) and (\ref{posdef4}), we find (reducing the
overall scaling factor)
\begin{equation}\label{posdef5}
m(x)-m(x_0)=\alpha x^2(nx-v)|^x_{x_0}\ ,
\end{equation}
or we have
\begin{equation}\label{posdef6}
m(x)=\alpha x^2(nx-v)+\mbox{constant.}
\end{equation}
However, since equation (\ref{basic1}) only takes care of regions
where $r>0$ while the mass at the origin $r=0$ is not included, this
constant is not arbitrary. A difference in this constant means a
difference in the total enclosed mass that changes with time in a
scaling of $t^{3n-s}$. The equation that includes the mass at $r=0$
is equation (\ref{basic2}), and this fixes the constant in equation
(\ref{posdef6}) to be zero. Thus, in order to introduce the total
enclosed mass into the radial momentum equation as a self-similar
variable, $nx-v>0$ is required for $3n-s>0$. Otherwise for $3n-s<0$,
we should require $nx-v<0$ accordingly.
%{\bf A reverse shock possible this way?}

\section[]{Determination of the Magnetosonic Critical\\ \quad Curve}\label{cridet}
%\section[]{Partial Derivatives of $A$, $V$ and $X$}\label{partial}

The magnetosonic critical curve of equation (\ref{transform8}) is
defined as the curve along which both the numerators and
denominators \{ $A(x,\alpha,v)$, $X(x,\alpha,v)$ and $V(x,\alpha,v)$
\} of equation (\ref{transform8}) vanish. In fact, only two of these
three vanishing conditions are independent and we shall simply use
the intersection of the two surfaces $A(x,\alpha,v)=0$ and
$X(x,\alpha,v)=0$ to determine the magnetosonic critical curve.

In order to obtain the magnetosonic critical curve numerically, we
denote $\theta\equiv nx-v$ and $\theta>0$ is required for a positive
total enclosed mass (see Appendix \ref{posdef}). We readily obtain
from $A(x,\alpha,v)=X(x,\alpha,v)=0$ the following two equations,
namely
\[
\alpha\!\!=\!\!\bigg\{\!\!q\theta^2\!\!
+\!\!\bigg(\frac{2-n}{3n-2}\!+\!\frac{q}{2}\bigg)\!\!
\bigg[(n-1)(nx+\theta)\theta\! -\!\frac{2\theta^3}x\bigg]\!\bigg\}
\]
\begin{equation}\label{cridet1}
\qquad\bigg/\bigg[qhx^2-\bigg(\frac{2-n}{3n-2}+\frac{q}{2}\bigg)
\bigg(2hx\theta+\frac{\theta^2}{3n-2}\bigg)\bigg]
\end{equation}
%The above version should be equivalent to the following
%commented out version. April 17, 2007 Tuesday
%\[
%\alpha\!\!=\!\!\bigg\{\!\!(3n-2)q\theta^2\!\!
%+\!\!\bigg[2-n\!+\!\frac{3n\!-\!2}2q\bigg]\!\!
%\bigg[(n-1)(nx+\theta)\theta\!
%-\!\frac{2\theta^3}x\bigg]\!\bigg\}
%\]
%\begin{equation}\label{cridet1}
%\times\bigg[(3n-2)qhx^2-\bigg(2-n+\frac{3n-2}2q\bigg)
%\bigg(2hx\theta+\frac{\theta^2}{3n-2}\bigg)\bigg]^{-1}\ ,
%\end{equation}
and
\begin{equation}\label{cridet2}
\bigg(2-n+\frac{3n-2}2q\bigg)\alpha^{1-n+3nq/2}x^{2q}\theta^q
+h\alpha x^2=\theta^2\ .
\end{equation}
A direct substitution of equation (\ref{cridet1}) into equation
(\ref{cridet2}) to replace $\alpha$ yields a nonlinear equation for
$x$ and $\theta$ which leads to a curve $\theta=\theta(x)$ for the
magnetosonic critical curve. Using this curve, the definition of
$\theta$ and equation (\ref{cridet1}), the magnetosonic critical
curve is then determined in the form of $(x,\alpha,v)$. In our
analysis, the relation $\theta=\theta(x)$ is determined numerically
for a set of three specified parameters $\{ n, q, h\}$; the
definition of $q$ contains the polytropic index $\gamma$ parameter.

\section[]{Eigensolutions crossing the
Magnetosonic\\ \quad Critical Curve}\label{eigsol}

%\section[]{Second-Order Derivatives of $\alpha(x)$
%and \lowercase{$\mathit{v(x)}$}}\label{second}

Using the l'H$\hat{\mbox{o}}$spital rule and with a superscript
prime `$\prime$' indicating a differentiation with respect to $x$,
we obtain the following differential relation along the magnetosonic
critical curve for $v'$ and $\alpha'$ from equation
$X(x,\alpha,v)v'=V(x,\alpha,v)$, namely
\[
[f_1(x,\alpha,v)\alpha'+f_2(x,\alpha,v)v'+f_3(x,\alpha,v)]v'
\]
\begin{equation}\label{eigsol1}
\qquad=f_4(x,\alpha,v)\alpha'+f_5(x,\alpha,v)v'+f_6(x,\alpha,v)\ ,
\end{equation}
where functional coefficients $f_i(x,\alpha,v)$ with $i=1,\cdots, 6$
are defined by
%{\bf You may want to introduce $\theta\equiv nx-v$ to
%simplify the following crowded expressions. }{\it to YQL: done}
\[
f_1(x,\alpha,v)\equiv hx^2+\bigg(2-n+\frac{3n-2}2q\bigg)
\bigg(1-n+\frac{3nq}{2}\bigg)
\]
\[
\qquad\qquad\qquad \times\alpha^{-n+3nq/2}x^{2q}\theta^q\ ,
\]
\[
f_2(x,\alpha,v)\equiv 2\theta-q\bigg(2-n+\frac{3n-2}2q\bigg)
\]
\[
\qquad\qquad\qquad\qquad \times\alpha^{1-n+3nq/2}x^{2q}\theta^{q-1}\
,
\]
\[
f_3(x,\alpha,v)\equiv 2h\alpha x-2n\theta
+2q\bigg(2-n+\frac{3n-2}2q\bigg)
\]
\[
\quad\times\alpha^{1-n+3nq/2}x^{2q-1}\theta^q
+nq\bigg(2-n+\frac{3n-2}2q\bigg)
\]
\[
\quad\times\alpha^{1-n+3nq/2}x^{2q}\theta^{q-1}\ ,
\]
\[
f_4(x,\alpha,v)\equiv 2(1-n)hx^2-\frac{\theta^2}{(3n-2)}
\]
\[
\qquad+[2(1-n)(2-n)-n(3n-2)q](1-n+3nq/2)
\]
\[
\qquad\times\alpha^{-n+3nq/2}x^{2q}\theta^q+[2(2-n)+(3n-2)q]
\]
\[
\qquad\times(1-n+3nq/2)\alpha^{-n+3nq/2}x^{2q-1}\theta^{q+1}\ ,
\]
\[
f_5(x,\alpha,v)\equiv (n-1)(nx-2\theta)+\frac{2\alpha\theta}{(3n-2)}
\]
\[
\quad-q[2(1\!-\!n)\!(2\!-\!n)\!-\!n(3n\!-\!2)q]\alpha^{1\!-\!n\!+\!3nq\!/\!2}x^{2q}\theta^{q\!-\!1}
\]
\[
\quad-\!(q\!+\!1)[2(2\!-\!n)\!+\!(3n\!-\!2)q]
\alpha^{1\!-\!n\!+\!3nq\!/\!2}x^{2q\!-\!1}\theta^q\
,
\]
\[
f_6(x,\alpha,v)\equiv 4(1-n)h\alpha x-2n^2(n-1)x+n(n-1)v
\]
\[
\quad+2n(n-1)\theta-{2n\alpha\theta}/{(3n-2)}
\]
\[
\quad+2q[2(1\!-\!n)\!(2\!-\!n)\!-\!n(3n\!-\!2)q]
\alpha^{1\!-\!n\!+\!3nq\!/\!2}x^{2q\!-\!1}\theta^q
\]
\[
\quad+nq[2(1\!-\!n)\!(2\!-\!n)\!-\!n(3n\!-\!2)q]
\alpha^{1\!-\!n\!+3nq\!/\!2}x^{2q}\theta^{q\!-\!1}
\]
\[
\quad+(2q\!-\!1)[2(2\!-\!n)\!+\!(3n\!-\!2)q]
\alpha^{1\!-\!n\!+\!3nq\!/\!2}x^{2q\!-\!2}\theta^{q\!+\!1}
\]
\begin{equation}\label{eigsol2}
\quad+n(q\!+\!1)[2(2\!-\!n)\!+\!(3n\!-\!2)q]
\alpha^{1\!-\!n\!+\!3nq\!/\!2}x^{2q\!-\!1}\theta^q\ ,
\end{equation}
[see equations (\ref{transform8}) and (\ref{transform9})]. Together
with the differential relation
\begin{equation}\label{eigsol3}
\alpha'=\frac{\alpha v'-2(x-v)\alpha/x}{(nx-v)}\ ,
\end{equation}
we obtain an algebraic quadratic equation for $v'\equiv dv/dx$ at
the magnetosonic critical curve, namely
\[
\bigg[ \frac\alpha{(nx-v)}f_1(x,\alpha,v)+f_2(x,\alpha,v)
\bigg](v')^2
\]
\[
\qquad
+\bigg[f_3(x,\alpha,v)-\frac{2(x-v)\alpha}{x(nx-v)}f_1(x,\alpha,v)
\]
\[
\quad\qquad
-\frac\alpha{(nx-v)}f_4(x,\alpha,v)-f_5(x,\alpha,v)\bigg]v'
\]
\begin{equation}\label{eigsol4}
\quad\quad\qquad
+\frac{2(x-v)\alpha}{x(nx-v)}f_4(x,\alpha,v)-f_6(x,\alpha,v)=0\ .
\end{equation}
The two real roots $v'$ of quadratic equation (\ref{eigsol4}) on the
magnetosonic critical curve represents two eigensolutions which can
go across the magnetosonic critical curve smoothly.

\section[]{Solution to MHD Shock Conditions}\label{shocksolve}

%\section[]{A Proof of $D_3=0$}\label{proofd3}

In reference to MHD shock jump equations
(\ref{shock6})$-$(\ref{shock8}), we may denote\footnote{Please note
that in defining similar new variables above equation (46) of Lou \&
Wang (2006), there was a typo in this regard.} $\Gamma_i\equiv
n-v_i/x_i$ and rearrange these MHD shock equations into the
following equivalent form
\[
\alpha_1\Gamma_1=\alpha_2\Gamma_2\ ,
\]
\[
\alpha_1^{2-n+3nq/2}x_1^{3q-2}\Gamma_1^q
+\alpha_1\Gamma_1^2+\frac{h\alpha_1^2}{2}
\]
\[
\qquad\quad=\alpha_2^{2-n+3nq/2}x_2^{3q-2}\Gamma_2^q
+\alpha_2\Gamma_2^2+\frac{h\alpha_2^2}{2}\ ,
\]
\[
\frac{2\gamma}{(\gamma-1)}\alpha_1^{1-n+3nq/2}x_1^{3q-2}
\Gamma_1^q+\Gamma_1^2+2h\alpha_1
\]
\begin{equation}\label{ssolve1}
\qquad\quad
=\frac{2\gamma}{(\gamma-1)}\alpha_2^{1-n+3nq/2}x_2^{3q-2}
\Gamma_2^q+\Gamma_2^2+2h\alpha_2\ .
\end{equation}
Using the first mass conservation equation in (\ref{ssolve1}), the
downstream reduced density $\alpha_2$ can be immediately replaced to
obtain
\[
\alpha_1^{2-n+3nq/2}x_1^{3q-2}\Gamma_1^q+\alpha_1\Gamma_1^2
+\frac{h\alpha_1^2}{2}
\]
\[
\quad=\frac{\alpha_1^{2-n+3nq/2}\Gamma_1^{2-n+3nq/2}}
{\Gamma_2^{2-n+(3n-2)q/2}}x_2^{3q-2}
+\alpha_1\Gamma_1\Gamma_2+\frac{h}{2}
\frac{\alpha_1^2\Gamma_1^2}{\Gamma_2^2}\ ,
\]
\[
\frac{2\gamma}{(\gamma-1)}\alpha_1^{1-n+3nq/2}
x_1^{3q-2}\Gamma_1^q+\Gamma_1^2+2h\alpha_1
\]
\[
\qquad\quad=\frac{2\gamma}{(\gamma-1)}\frac{\alpha_1^{1-n+3nq/2}
\Gamma_1^{1-n+3nq/2}}{\Gamma_2^{1-n+(3n-2)q/2}}x_2^{3q-2}
\]
\begin{equation}\label{ssolve2}
\qquad\qquad\qquad\qquad
+\Gamma_2^2+2h\frac{\alpha_1\Gamma_1}{\Gamma_2}\ .
\end{equation}
Eliminating $x_2^{(3q-2)}$ term in both equations of
(\ref{ssolve2}), we obtain a cubic equation for $\Gamma_2$ of the
downstream side (subscript 2) in terms of upstream variables
(subscript 1)
\[
\frac{(\gamma+1)}{2\gamma}\Gamma_2^3
-\bigg(\alpha_1^{1-n+3nq/2}x_1^{3q-2}\Gamma_1^{q-1}
+\Gamma_1+\frac{h\alpha_1}{2\Gamma_1}\bigg)\Gamma_2^2
\]
\[
\quad+\bigg(\alpha_1^{1-n+3nq/2}x_1^{3q-2}\Gamma_1^q
+\frac{\gamma-1}{2\gamma}\Gamma_1^2
+\frac{\gamma-1}{\gamma}h\alpha_1\bigg)\Gamma_2
\]
\begin{equation}\label{ssolve3}
\quad+\frac{(2-\gamma)}{2\gamma}h\alpha_1\Gamma_1=0\ .
\end{equation}
Finally, because $\Gamma_2=\Gamma_1$ is a trivial and unphysical
solution of equation (\ref{ssolve3}) for an MHD shock, we then
obtain a quadratic equation of $\Gamma_2$ for possible physical MHD
shock solutions
\[
\frac{(\gamma+1)}{2\gamma}\Gamma_2^2
-\bigg(\alpha_1^{1-n+3nq/2}x_1^{3q-2}
\Gamma_1^{q-1}+\frac{\gamma-1}{2\gamma}\Gamma_1
+\frac{h\alpha_1}{2\Gamma_1}\bigg)\Gamma_2
\]
\begin{equation}\label{ssolve4}
\qquad\qquad\qquad -\frac{(2-\gamma)}{2\gamma}h\alpha_1=0\ ,
\end{equation}
and the positive $\Gamma_2$ root of this quadratic equation
represents a proper MHD shock solution.

%{\bf Complex $x_i$?} {\bf YQL added the following paragraph.
%Please check}{\it to YQL: OK.}

For $\Gamma_1>0$ and $1<\gamma<2$, quadratic equation
(\ref{ssolve4}) always has two real roots for $\Gamma_2$ with the
plus-sign root being positive and the minus-sign root being
negative. The first relation of equation (\ref{ssolve1}) gives the
corresponding $\alpha_2$. One can usually determine a sensible $x_2$
and thus $v_2$ and $\lambda$ values. Occasionally, we encounter the
unrealizable situation of a complex $x_2$ caused by the presence of
a random magnetic field (i.e., $h\neq 0$). For $h=0$, the situation
of a complex $x_2$ would not arise. This suggests that self-similar
MHD shocks cannot form within a certain radius and time for a given
set of parameters.

%\section[]{Determination of Magnetosonic
%Critical Curves}\label{infocritical}

%\section[]{MHD Eigensolution Behaviours in the Vicinity
%of the Magnetosonic Critical Curve}\label{vicinity}

%\section[]{Asymptotic Behaviours Approaching
%the Quasi Magnetostatic Solution}\label{asystat}

\end{document}